\documentclass[twocolumn]{aastex63}
\usepackage{natbib}
\usepackage{amsmath}
\usepackage{floatrow}
\usepackage{color}
\usepackage{graphicx}
\submitjournal{AJ}

\shorttitle{ExoTETHyS II}
\shortauthors{Morello et al.}
\begin{document}

\title{Phase curve pollution of exoplanet transmission spectra}

\correspondingauthor{Giuseppe Morello}
\email{gmorello@iac.es}

\author[0000-0002-4262-5661]{Giuseppe Morello}
\affiliation{AIM, CEA, CNRS, Universit\'e Paris-Saclay, Universit\'e de Paris, F-91191 Gif-sur-Yvette, France}
\affiliation{Dept. of Physics \& Astronomy, University College London, Gower Street, WC1E 6BT London, UK}
\affiliation{INAF--Osservatorio Astronomico di Palermo, Piazza del Parlamento 1, I-90134 Palermo, Italy}

\author[0000-0001-6880-5356]{Tiziano Zingales}
\affiliation{Laboratoire d'astrophysique de Bordeaux, Univ. Bordeaux, CNRS, B18N, all\'{e}e Geoffroy Saint-Hilaire, F-33615 Pessac, France}
\affiliation{INAF--Osservatorio Astronomico di Palermo, Piazza del Parlamento 1, I-90134 Palermo, Italy}

\author[0000-0003-0523-7683]{Marine Martin-Lagarde}
\affiliation{AIM, CEA, CNRS, Universit\'e Paris-Saclay, Universit\'e de Paris, F-91191 Gif-sur-Yvette, France}

\author{Ren{\'e} Gastaud}
\affiliation{AIM, CEA, CNRS, Universit\'e Paris-Saclay, Universit\'e de Paris, F-91191 Gif-sur-Yvette, France}

\author{Pierre-Olivier Lagage}
\affiliation{AIM, CEA, CNRS, Universit\'e Paris-Saclay, Universit\'e de Paris, F-91191 Gif-sur-Yvette, France}



\begin{abstract}
The occurrence of a planet transiting in front of its host star offers the opportunity to observe the planet's atmosphere filtering starlight.
The fraction of occulted stellar flux is roughly proportional to the optically thick area of the planet, the extent of which depends on the opacity of the planet's gaseous envelope at the observed wavelengths.
Chemical species, haze, and clouds are now routinely detected in exoplanet atmospheres through rather small features in transmission spectra, i.e., collections of planet-to-star area ratios across multiple spectral bins and/or photometric bands.
Technological advances have led to a shrinking of the error bars down to a few tens of parts per million (ppm) per spectral point for the brightest targets.
The upcoming James Webb Space Telescope (JWST) is anticipated to deliver transmission spectra with precision down to 10 ppm.
The increasing precision of measurements requires a reassessment of the approximations hitherto adopted in astrophysical models, including transit light curve models.
Recently, it has been shown that neglecting the planet's thermal emission can introduce significant biases in the transit depth measured with the JWST/Mid-InfraRed Instrument, integrated between 5 and 12 $\mu$m.
In this paper, we take a step forward by analyzing the effects of the approximation on transmission spectra over the 0.6-12 $\mu$m wavelength range covered by various JWST instruments. We present open source software to predict the spectral bias, showing that, if not corrected, it may affect the inferred molecular abundances and thermal structure of some exoplanet atmospheres.
\end{abstract}
%WARNING: maybe 249 words in the abstract (max is 250).



\section{Introduction} \label{sec:intro}
The transit method has been the most successful for the detection and characterization of exoplanets to date. The CoRoT \citep{auvergne2009}, Kepler \citep{borucki2010}, and K2 \citep{howell2014} space missions have discovered nearly 3\,000 transiting exoplanets. Ground-based surveys, such as HATNet \citep{bakos2004}, HATSouth \citep{bakos2013}, WASP \citep{pollacco2006}, and TRAPPIST \citep{gillon2016}, have added hundreds of transiting exoplanets to the current census.
%The \emph{Transiting Exoplanet Survey Satellite} (\emph{TESS}, \citealp{ricker2016}), launched in April 2018, is expected to discover more than 4\,000 exoplanet transiting around nearby stars, and about 10\,000 around fainter stars, during its 2-year primary mission time \citep{barclay2018}. The \emph{PLAnetary Transits and Oscillations of Stars} (\emph{PLATO}, \citealp{rauer2014}), will be able to find small transiting planets in the habitable zones of solar-like stars \citep{rauer2018}.

Transit photometry can reveal the size of the planet related to its host star and the orbital inclination based on the depth and duration of the drop in flux that occurs as the planet transits between the star and the observer for a fraction of its orbit. If combined with stellar spectra and radial velocity measurements, transit observations inform us about the mass, radius, mean density, and equilibrium temperature of exoplanets.

Transit spectroscopy provides insights into the atmospheres of exoplanets based on precise measurements of the transit depth at multiple wavelengths. Recently, the Wide Field Camera 3 (WFC3) on board the Hubble Space Telescope (HST) delivered dozens of low-resolution transmission spectra in the 0.8-1.7 $\mu$m range \citep{iyer2016,fu2017,fisher2018,tsiaras2018}. The HST/WFC3 spectra have revealed the presence of H$_2$O vapor in the atmosphere of several exoplanets down to super-Earth mass \citep{deming2013,damiano2017,benneke2019b,tsiaras2019,edwards2020} and hints of TiO \citep{evans2016}, He \citep{mansfield2018,spake2018}, FeH \citep{skaf2020}, H$^-$ \citep{pluriel2020}, and AlO \citep{chubb2020b}.
Other HST and ground-based spectra in the UV and visible wavelengths have enabled detection of a long list of ions, atoms, haze, and clouds in some exoplanet atmospheres \citep{charbonneau2002,vidal-madjar2003,vidal-madjar2004,fossati2010,linsky2010,parviainen2016,chen2017,chen2018}.
The Spitzer/InfraRed Array Camera (IRAC) broadband photometry at 3.6, 4.5, 5.8, and 8.0 $\mu$m has suggested the possible presence of carbon molecules \citep{beaulieu2011,knutson2011,morello2015}.
Combined spectra from different space- and/or ground-based instruments covering UV to infrared wavelengths have also been analyzed in the literature \citep{danielski2014,sing2016,barstow2017,mancini2019,pinhas2019,luque2020}.

Typical error bars in transit spectra can be a few tens to hundreds of parts per million (ppm). 
The upcoming James Webb Space Telescope (JWST) will provide infrared spectra with unprecedented precision down to $\sim$10 ppm for some targets \citep{beichman2014}. 
The Ariel mission aims to perform similarly high-precision transit spectroscopy over a statistical sample of $\sim$1\,000 exoplanet atmospheres \citep{tinetti2018}.
In parallel with the improvements of instruments and observational techniques, it is necessary to develop more detailed models to adequately represent astrophysical phenomena, such as stellar limb-darkening, and magnetic activity \citep{micela2015,scandariato2015,morello2017,morello2018,sarkar2018}.

Here we discuss the impact of the ``dark planet'' approximation in transit spectroscopy. As the planet is orders of magnitude fainter than the host star, its flux contribution is usually neglected when modeling the transit light curves \citep{mandel2002}.
\cite{kipping2010} pointed out that the measured transit depth tends to be underestimated due to the planetary flux acting as a self-blend. They evaluated that the bias in transit depth could be down to -100 ppm for some hot Jupiters observed in the thermal infrared.
\cite{martin-lagarde2020} showed that the transit depth can also be overestimated due to the variation of the emergent planetary flux with the orbital phase, the so-called ``phase-blend'' effect. They predicted the total bias in transit depth for a sample of exoplanets observed with the JWST/Mid-InfraRed Instrument (MIRI) to be between -100 and +550 ppm, depending on the system parameters and exoplanet atmospheric properties. They also noted that, unlike the self-blend, the phase-blend effect is degenerate with instrumental systematics that can be detrended from the data.

In this paper, we explore how self- and phase-blend biases can alter the transmission spectra of exoplanet atmospheres and consequently affect scientific inferences.

\paragraph{Structure of the paper}
Section \ref{sec:boats} describes the newly released \texttt{ExoTETHyS.BOATS} subpackage, with particular emphasis on the procedure to compute the self- and phase-blend biases. Section \ref{sec:spectral_bias} discusses some examples of spectral biases. Section \ref{sec:biased_retrievals} presents a study of simulated spectra obtained with JWST spectrographs. Section \ref{sec:jwst_gto_ers} discusses the potential impact of self- and phase-blend bias in future observations with JWST, focusing on the targets that are already scheduled. Section \ref{sec:conclusions} summarizes the conclusions of our study.

\section{The \texttt{ExoTETHyS.BOATS} subpackage} \label{sec:boats}
We present here a new tool designed to estimate the Bias in the Occultation Analysis of Transiting Systems (BOATS). The BOATS subpackage\footnote{ \url{https://github.com/ucl-exoplanets/ExoTETHyS/blob/master/exotethys/boats.py}} is part of \texttt{ExoTETHyS}, an open-source software that collects a set of tools for modeling exoplanetary transits and their host stars \citep{morello2020, morello2020joss}. All of the code within \texttt{ExoTETHyS} is written in python 3.

The BOATS subpackage contains several functions designed to obtain the spectra of astrophysical objects, to perform specific operations on these spectra (e.g., conversions, integration over instrumental passbands, time intervals and/or areas, and composition of spectra), and to compute some properties of transiting exoplanets (e.g., transit duration, atmospheric temperatures, albedo, and circulation efficiency).

In this paper, we describe how to use the BOATS subpackage to obtain estimates of the potential self- and phase-blend biases in observations of exoplanetary transits.

\subsection{Input and Output}
The procedure requires a configuration file to perform the full set of predetermined calculations for a given exoplanet transit or eclipse. The user can choose between several model grids with precalculated stellar spectra, use a blackbody spectrum, or read the stellar spectrum from a file. The choice for exoplanets is between blackbody and user file spectra to model their dayside and nightside emission. In the case of spectra read from user files, it is necessary to specify whether these spectra are to be rescaled by the distance and radius of the source to obtain the observed flux. The input planetary spectra can be given in absolute units or normalized with respect to the stellar spectrum, such as those typically obtained by running forward models of the secondary eclipses with NEMESIS \citep{irwin2008}, CHIMERA \citep{line2013}, and Tau-REx II \citep{waldmann2015} and III \citep{al-refaie2019}.

The user must request one or more instrumental passbands. Some passbands are built into the \texttt{ExoTETHyS} package through text files that report the photon-to-electron conversion efficiencies (PCEs) at given wavelengths. User files with the same format can also be accepted. Optionally, the passbands can be split into multiple wavelength bins, as is usual in exoplanet transit spectroscopy.

Other mandatory input information includes the area of the telescope ($\mathcal{A}_{\text{tel}}$), the duration of the simulated observation ($\Delta t_{\text{obs}}$), and a set of parameters of the star--planet system. The distance of the system from the Earth ($d$) and the telescope area can be used to rescale the template spectra in order to match the photon count rates received by the detector. We anticipate here that rescaling by $d$ and $\mathcal{A}_{\text{tel}}$ does not affect the transit depth biases, only the nominal error bars.

The input stellar parameters are the effective temperature ($T_{*,\text{eff}}$), surface gravity ($\log{g_*}$), metallicity ($[$M/H$]_*$), and radius ($R_*$). The stellar models in the grids are identified by $T_{*,\text{eff}}$, $\log{g_*}$, and $[$M/H$]_*$. The default values of $\log{g_*}=$4.5 and $[$M/H$]_*=$0.0 are implemented if these parameters are missing in the configuration file. The blackbody spectrum is defined by $T_{*,\text{eff}}$ alone. If the stellar spectrum is obtained from a user file, none of these three parameters is necessary, unless $T_{*,\text{eff}}$ is required to calculate the planetary spectra. Here $R_*$ is always mandatory.

The orbital period ($P$), semimajor axis ($a$), inclination ($i$), and planet radius ($R_p$) are needed to determine the geometry of the transit. The planet's emission is modeled through two spectra from the dayside and nightside hemispheres, either blackbody or user-supplied spectra. The blackbody spectra can be defined by directly setting the dayside and nightside temperatures ($T_{\text{day}}$, $T_{\text{night}}$), or by means of the bond albedo ($A_b$) and circulation efficiency ($\varepsilon$).

Each physical parameter, except for temperatures and dimensionless parameters, is defined through two keywords in the configuration file to give its numerical value and the corresponding unit. The \texttt{astropy.units} package is used to interpret the input units and to handle operations between physical quantities. Additionally, $R_p$ and $a$ can be expressed in units of $R_*$, and $\Delta t_{\text{obs}}$ can be relative to the total transit duration ($T_{1-4}$).

The output is a text or pickle file containing estimates of transit depth biases and error bars for the requested exoplanet systems and passbands.

\subsection{Flux conversions}

The stellar model atmosphere grids consist of one file for each triple of stellar parameters ($T_{*,\text{eff}}$, $\log{g_*}$, $[$M/H$]_*$), containing the emergent spectrum at the star surface in units of erg\,cm$^{-2}$\,s$^{-1}$\,\AA$^{-1}$.
The same files also include spatially resolved intensities, which are used by other subpackages to model limb-darkening.
The BOATS algorithm first calculates the spectral photon rates collected by the telescope in units of photon\,s$^{-1}$\,\AA$^{-1}$. In order to do so, the stellar and planetary spectra are multiplied by $( R / d )^2 \mathcal{A}_{\text{tel}}$, where $R$ is the object's radius. The geometric dilution factor  $( R / d )^2$ can be set to 1 for a user file spectrum by selecting the no-rescale flux option.

The next step is to compute the electron rates of detectors. This is done by multiplying the photon rates by the PCE and integrating on the passband or wavelength bin.
In the following subsections, we will refer to fluxes (symbol $F$) as equivalent to electron rates in e$^-$\,s$^{-1}$, calculated as described here.

\subsection{Exoplanet phase curve model}
\label{ssec:pc_model}

The BOATS calculations rely on a toy model to describe the phase curve modulations. The observed planet flux is
\begin{equation}
\label{eqn:Fp_psi}
F_p = \psi F_{\text{day}} + (1-\psi) F_{\text{night}} ,
\end{equation}
where $F_{\text{day}}$ and $ F_{\text{night}}$ denote the emergent flux from the two hemispheres of the planet, and $\psi$ is the fraction of the area of the planet's projection showing the dayside. Other assumptions (valid for \texttt{ExoTETHyS} version 2.0.0) are that the planet lies on a circular, edge-on orbit, in synchronous rotation with zero obliquity, and the dayside is centered on the substellar point. Under these hypotheses, we have
\begin{equation}
\label{eqn:psi_phi}
\psi = \frac{1 - \cos{\varphi}}{2} ,
\end{equation}
where
\begin{equation}
\label{eqn:phi_definition}
\varphi = \frac{2 \pi}{P} t
\end{equation}
is the orbital phase angle, and $t$ is the time from mid-transit.
This toy model is identical to that adopted by the \texttt{ExoNoodle} software \citep{martin-lagarde2020, martin-lagarde2020joss}, but with one less degree of freedom relative to the centering of the dayside hemisphere with respect to the substellar point. It relies on the simplest hypotheses that enable order-of-magnitude estimates of the possible self- and phase-blend effects, solely based on the easiest-to-measure system parameters.
%We note that releasing some of these assumptions might reduce the phase curve variation due to a restricted range for $\psi$.

If Blackbody is the selected grid of planetary models, the two temperatures, $T_{\text{day}}$ and $T_{\text{night}}$, can also be determined from the atmospheric parameters $A_b$ and $\varepsilon$ \citep{cowan2011},
\begin{equation}
\begin{cases}
T_{\text{day}} = T_{\text{irr}}\left(1 - A_b\right)^{\frac{1}{4}} \left(\frac{2}{3} - \frac{5}{12} \varepsilon \right)^{\frac{1}{4}} \\
T_{\text{night}} = T_{\text{irr}}\left(1 - A_b\right)^{\frac{1}{4}} \left(\frac{\varepsilon}{4}  \right)^{\frac{1}{4}}
\end{cases},
\end{equation}
where the irradiation temperature is
\begin{equation}
T_{\text{irr}} = T_{*,\text{eff}} \sqrt{ \frac{R_*}{a} } .
\end{equation}

In general, the dayside flux is the sum of planetary intrinsic emission and reflected starlight,
\begin{equation}
F_{\text{day}} = F_{\text{day}}^{\text{emission}} + F_{\text{day}}^{\text{reflection}} ,
\end{equation}
where $F_{\text{day}}^{\text{emission}}$ is calculated from the corresponding blackbody or user spectrum, and
\begin{equation}
F_{\text{day}}^{\text{reflection}} = A_b \left ( \frac{R_p}{2a} \right )^2 F_* ,
\end{equation}
with $F_*$ being the stellar flux. The nightside flux consists of a pure planetary emission component.

\subsection{Phase-averaged fluxes}
The next important step is to calculate the mean flux of the planet during and out of the transit. 
Given the symmetry of the phase curve model described in section \ref{ssec:pc_model}, and assuming that the observation interval is centered on the transit, it is sufficient to consider the half observation after the mid-transit point. The transit event is delimited by the external contact points; the corresponding duration is \citep{seager2003}
\begin{equation}
T_{1-4} = \frac{P}{\pi} \arcsin{ \left ( \frac{ \sqrt{ \left ( 1 + \frac{R_p}{R_*} \right )^2 - \left ( \frac{a}{R_*} \cos{i} \right )^2 }}{ \frac{a}{R_*} \sin{i} } \right ) }.
\end{equation}
The phase angles corresponding to the end of the transit and of the observation are \begin{equation}
\frac{\Phi_{1-4}}{2} = \frac{\pi}{P} T_{1-4}, \, \text{and} \, \frac{\Delta \varphi_{\text{obs}}}{2} = \frac{\pi}{P} \Delta t_{\text{obs}}.
\end{equation}
By using Equations \ref{eqn:Fp_psi}-\ref{eqn:phi_definition}, the phase-averaged fluxes are
%\begin{eqnarray}
\begin{multline}
F_{p}^{\text{in}} = \frac{2}{\Phi_{1-4}} \int_{0}^{\frac{\Phi_{1-4}}{2}} F_p d \phi = \\
F_{\text{day}} \overline{\psi}^{\text{in}} + F_{\text{night}} \left ( 1 - \overline{\psi}^{\text{in}} \right ) , \, \text{and} 
\end{multline}
\begin{multline}
F_{p}^{\text{out}} = \frac{2}{ \Delta \varphi_{\text{obs}} - \Phi_{1-4} } \int_{\frac{\Phi_{1-4}}{2}}^{\frac{\Delta \varphi_{\text{obs}}}{2}} F_p d \phi =\\
F_{\text{day}} \overline{\psi}^{\text{out}} + F_{\text{night}} \left ( 1 - \overline{\psi}^{\text{out}} \right ) ,
\end{multline}
%\end{eqnarray}
with
\begin{eqnarray}
\overline{\psi}^{\text{in}} = \frac{1}{2} \left ( 1 - \frac{ \sin{ \left ( \frac{\Phi_{1-4}}{2} \right ) }}{ \frac{\Phi_{1-4}}{2} } \right ), \, \text{and} \\
\overline{\psi}^{\text{out}} = \frac{1}{2} \left ( 1 - \frac{ \sin{ \left ( \frac{\Delta \varphi_{\text{obs}}}{2} \right ) } - \sin{ \left ( \frac{\Phi_{1-4}}{2} \right ) } }{ \frac{\Delta \varphi_{\text{obs}}}{2} - \frac{\Phi_{1-4}}{2} } \right ).
\end{eqnarray}

\subsection{Transit depth bias and error bars}
The transit depth is defined as the planet-to-star projected area ratio,
\begin{equation}
p^2 = \left ( \frac{R_p}{R_*} \right )^2 .
\end{equation}
Finally, the self- and phase-blend effects are calculated as (\citealp{martin-lagarde2020}, see their Figure 2)
\begin{eqnarray}
\label{eqn:deltap2_selfblend}
\left ( \Delta p^2 \right )_{\text{self-blend}} = - \frac{ F_{p}^{\text{out}} }{ F_* + F_{p}^{\text{out}} } p^2, \, \text{and} \\
\label{eqn:deltap2_phaseblend}
\left ( \Delta p^2 \right )_{\text{phase-blend}} = \frac{ F_{p}^{\text{out}} -  F_{p}^{\text{in}} }{ F_* + F_{p}^{\text{out}} } ,
\end{eqnarray}
and the total bias in transit depth is
\begin{equation}
\Delta p^2 = \left ( \Delta p^2 \right )_{\text{self-blend}} + \left ( \Delta p^2 \right )_{\text{phase-blend}} .
\end{equation}
In the photon noise limit, the minimum error bar depends on the number of electrons produced in and out of the transit. Ignoring the ingress, egress, and limb-darkening effects, these numbers are
\begin{eqnarray}
N^{\text{in}} = \left ( F_* (1-p^2) + F_{p}^{\text{in}} \right ) T_{1-4} , \, \text{and} \\
N^{\text{out}} = \left ( F_* + F_{p}^{\text{out}} \right ) \left ( \Delta t_{\text{obs}} - T_{1-4} \right ) .
\end{eqnarray}
The nominal error bar is
\begin{equation}
\sigma_{p^2} = \left ( 1 - p^2 \right ) \sqrt{ \frac{1}{N^{\text{in}}} + \frac{1}{N^{\text{out}}} } .
\end{equation}
We note that the actual error bar obtained by fitting the light curve data can be larger because of the parameter correlations and additional noise components.

\section{Spectroscopic trends} \label{sec:spectral_bias}

\begin{figure*}[t!]
\includegraphics[width=\textwidth]{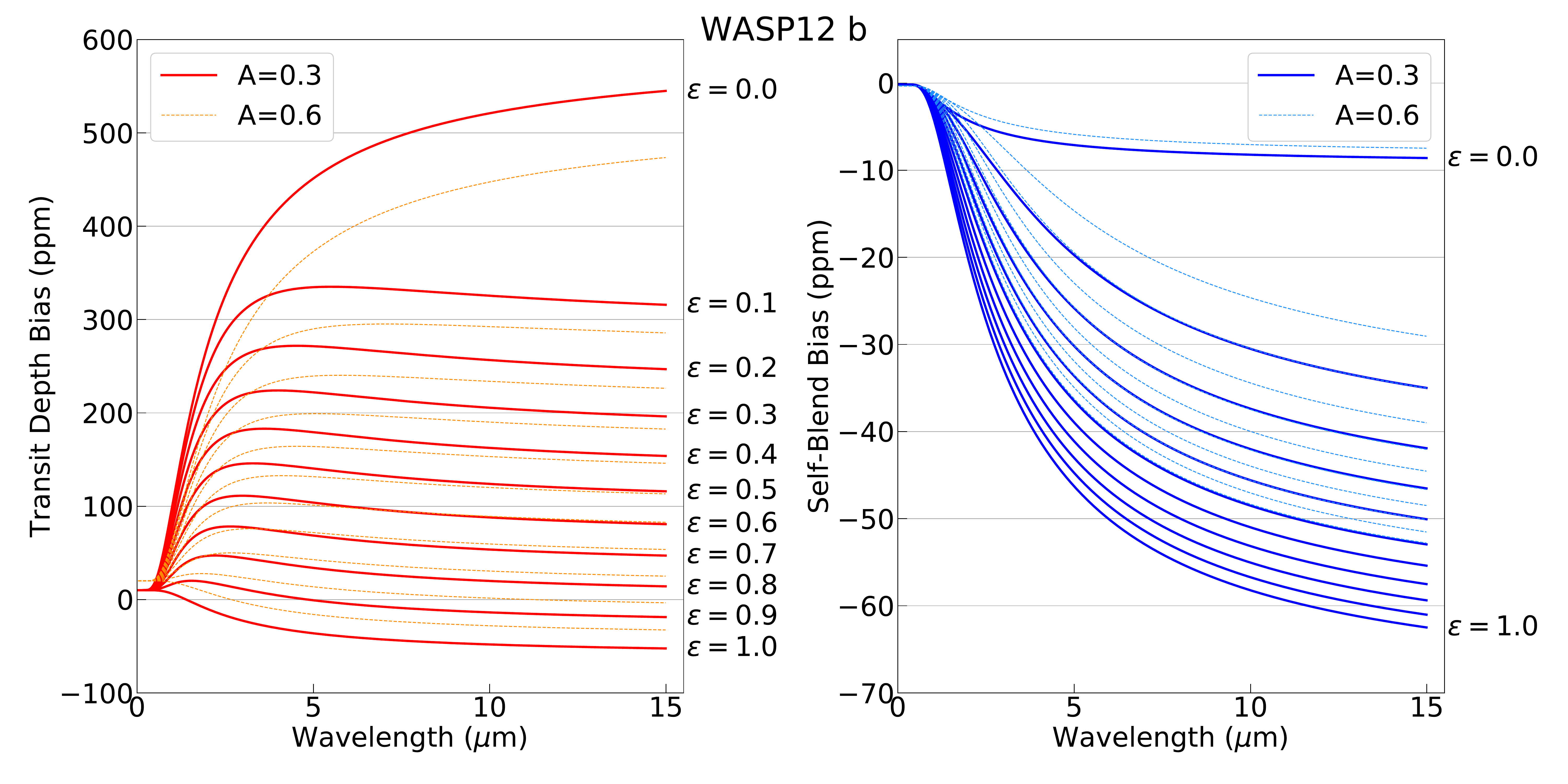}
\includegraphics[width=\textwidth]{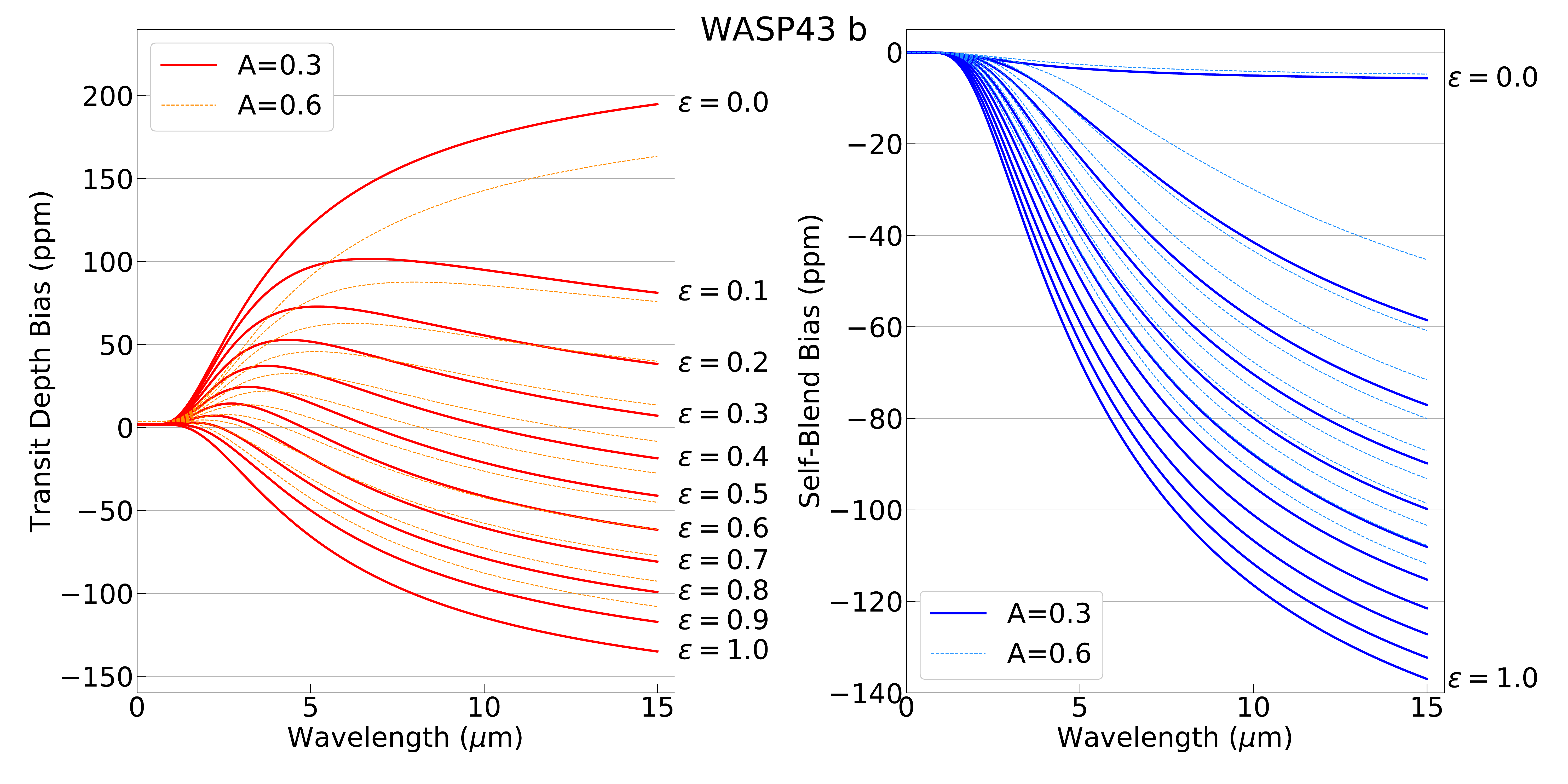}
\caption{Top left panel: total transit depth bias (in ppm) vs. wavelength for WASP~12~b. We assume pure blackbody emission from the exoplanet dayside and nightside hemispheres, different values of the atmospheric circulation efficiency ($0 \le \varepsilon \le 1$, step of 0.1), and bond albedo ($A_b=0.3, \, 0.6$). Top right panel: Analogous plot for the self-blend bias only. Bottom panels: Analogous plots for WASP~43~b. The self- and phase-blend bias increase in absolute value with wavelength as the star--planet contrast decreases. The self-blend (phase-blend) bias is always negative (positive), and it is larger with a higher (lower) circulation efficiency. The bond albedo acts as a scaling factor. The total bias is the sum of the two terms. The phase-blend alone is not shown, as it cannot be the only bias term.\label{fig:bb_bias_Ab_eps}}
\end{figure*}

\begin{figure*}[t!]
\includegraphics[width=0.49\textwidth]{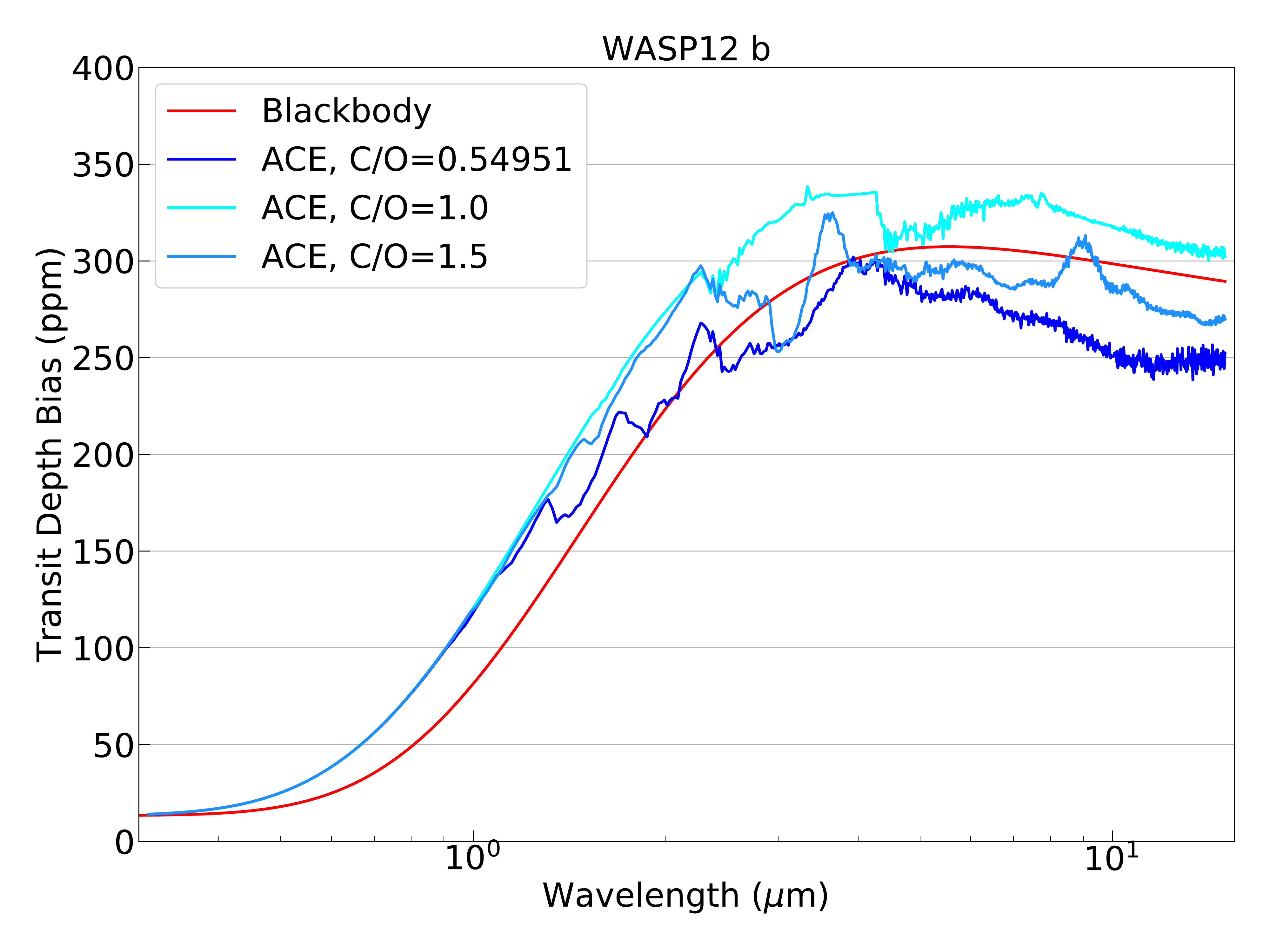}
\includegraphics[width=0.49\textwidth]{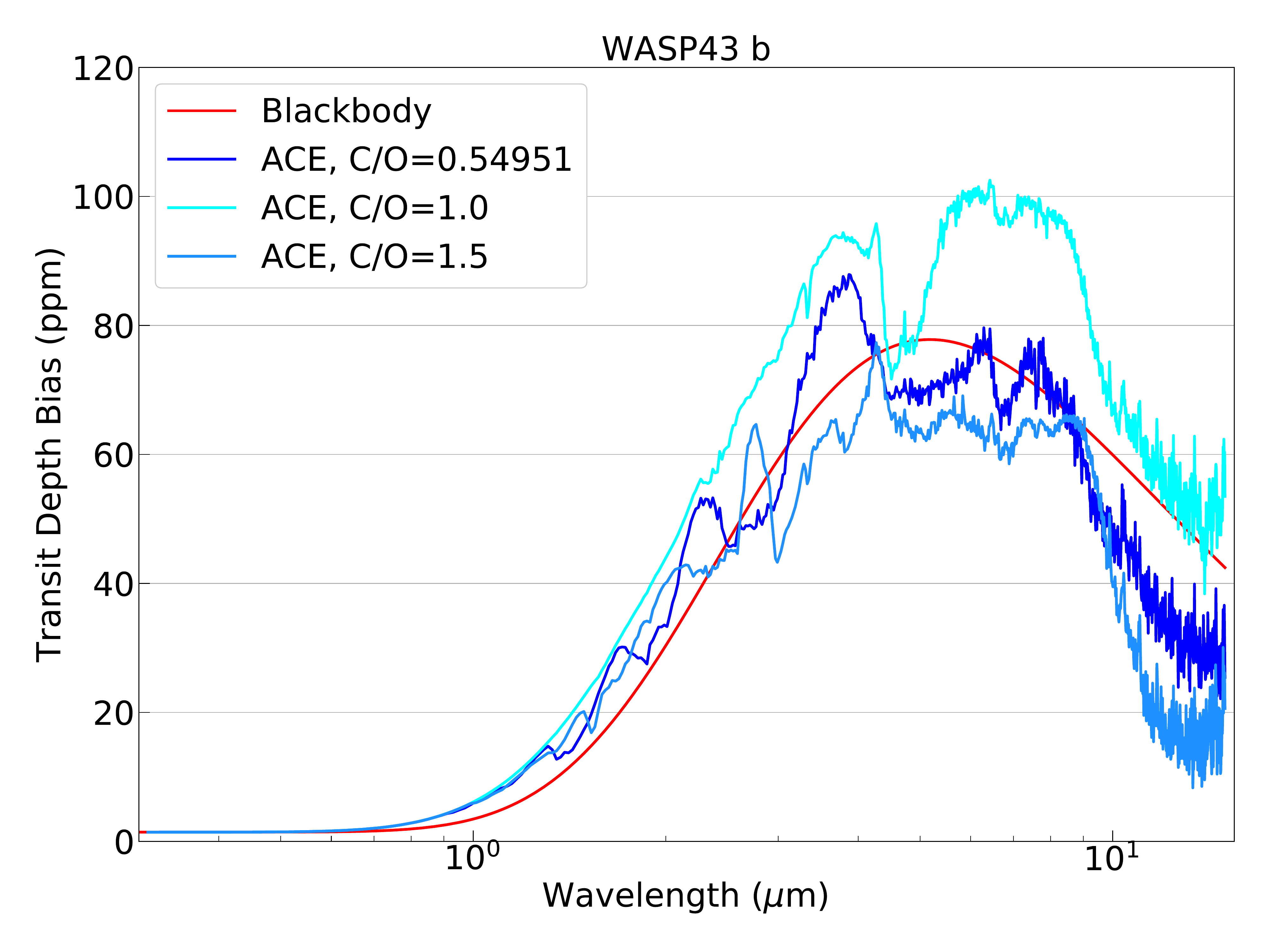}
\caption{Left panel: total transit depth bias (in ppm) vs. wavelength for WASP~12~b, assuming a pure blackbody (red), ACE with solar C/O ratio (blue), and C/O=1 (cyan) and 1.5 (dodger blue) emission spectra from the dayside and nightside hemispheres. The temperatures are fixed to the most likely values reported in Table \ref{tab:T_Ab_eps}. Right panel: Analogous plot for WASP~43~b. \label{fig:total_bias_bb_ace}}
\end{figure*}

\begin{deluxetable}{ccccccc}[h!]
\tablecaption{Nominal wavelength ranges and choice of bin sizes for the selected JWST observing modes. \label{tab:jwst_obs_modes}}
\tablecolumns{7}
%\tablenum{1}
\tablewidth{0pt}
\tablehead{
\colhead{Instrument} & \colhead{Range $\lambda$ ($\mu$m)} & \colhead{Bin $\Delta \lambda$ ($\mu$m)}
}
\startdata
NIRISS SOSS & 0.60--2.80 & 0.02--0.1\tablenotemark{a} \\
NIRSpec G235H & 1.66--3.17 & 0.03 \\
NIRSpec G395H & 2.87--5.27 & 0.04 \\
MIRI LRS & 5.0--12.0 & 0.1--0.2\tablenotemark{b} \\
\hline
\hline
\enddata
\tablenotetext{a}{NIRISS SOSS: $\Delta \lambda =0.02$ if $0.92<\lambda<1.98$, $\Delta \lambda =0.03$ if $0.80<\lambda<0.92$ or $1.98<\lambda<2.34$, $\Delta \lambda =0.04$ if $2.34<\lambda<2.50$, $\Delta \lambda =0.05$ if $0.70<\lambda<0.80$ or $2.50<\lambda<2.80$, and $\Delta \lambda =0.10$ if $0.60<\lambda<0.70$.}
\tablenotetext{b}{MIRI LRS: $\Delta \lambda =0.1$ if $5.0<\lambda<9.0$, and $\Delta \lambda =0.2$ if $9.0<\lambda<12.0$.}
\end{deluxetable}

It is evident from the Equations \ref{eqn:deltap2_selfblend} and \ref{eqn:deltap2_phaseblend} that the transit depth bias due to the dark planet approximation is wavelength-dependent, being a function of stellar and planetary fluxes. Consequently, it can alter the measurements of the transmission spectra of the exoplanet atmospheres.

Figure~\ref{fig:bb_bias_Ab_eps} shows the spectroscopic biases obtained for two systems with different values of bond albedo and circulation efficiency of the exoplanet atmospheres, assuming pure blackbody emissions. In all cases, the bias appears to be very small in the UV and visible, and it tends to be larger at wavelengths $\gtrsim$1 $\mu m$. The spectrum bias has a smaller amplitude with higher albedo due to the lower temperatures of the exoplanet atmosphere, but the overall trend is weakly dependent on the albedo.

\begin{figure}[h!]
\includegraphics[width=0.99\textwidth]{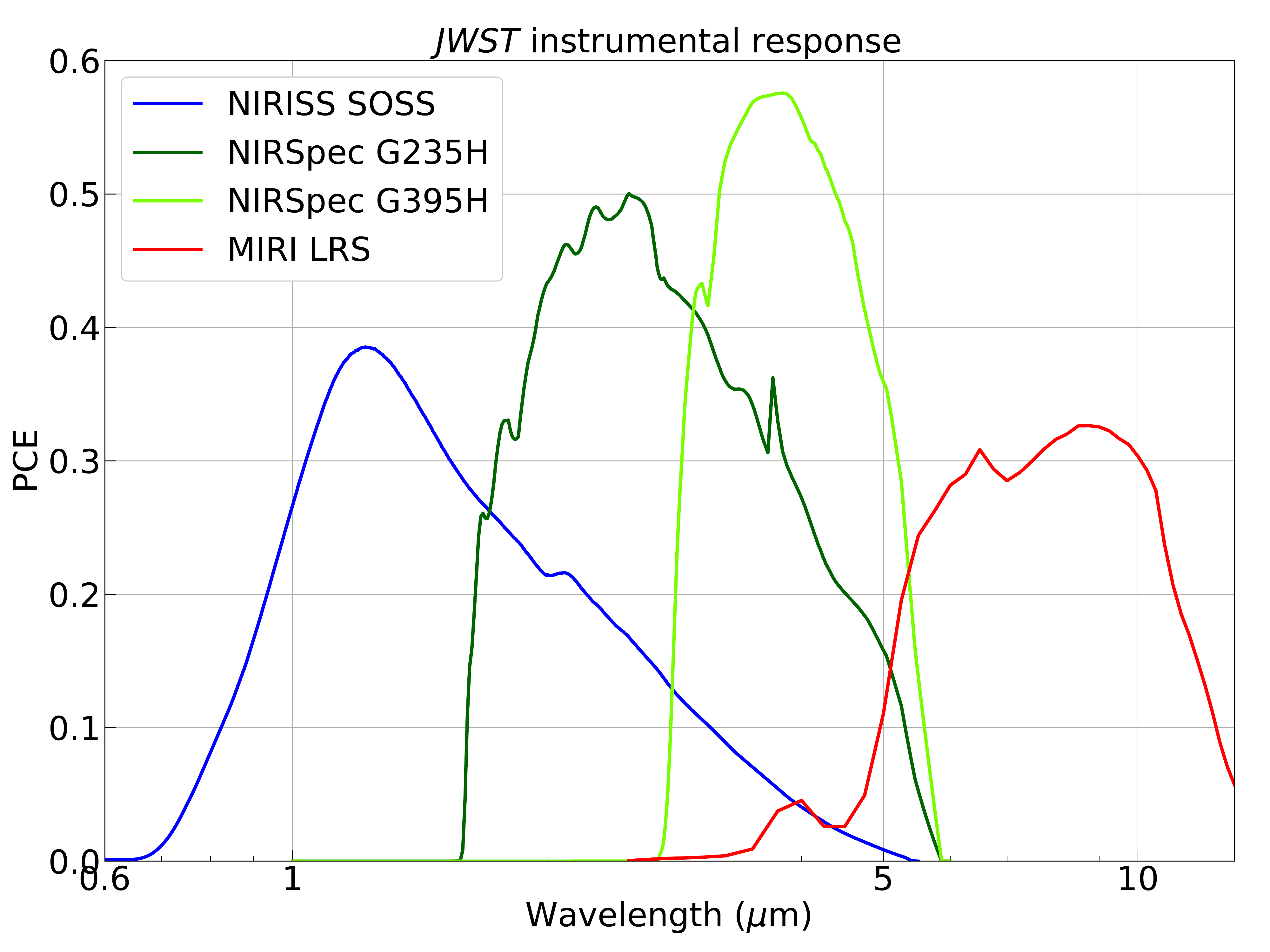}
\caption{The PCEs for the JWST NIRISS SOSS (blue), NIRSpec G235H (dark green), NIRSpec G395H (light green), and MIRI LRS (red) instrumental setups adopted in our simulations. \label{fig:PCEs}}
\end{figure}

\begin{deluxetable*}{ccccccccc}[t!]
\tablecaption{Adopted system parameters. \label{tab:system_parameters}}
\tablecolumns{9}
%\tablenum{1}
\tablewidth{0pt}
\tablehead{
\colhead{Planet} & \colhead{$R_p$ ($R_J$)} & \colhead{$a$ (au)} & \colhead{$i$ (deg)} & \colhead{$P$ (days)} & \colhead{$R_*$ ($R_\odot$)} & \colhead{$T_{*,\text{eff}}$ ($K$)} & \colhead{$d$ (pc)\tablenotemark{a}} & \colhead{References}
}
\startdata
WASP~12~b & 1.9 & 0.02339 & 83.37 & 1.0914203 & 1.657 & 6360 & 432.589 & \cite{collins2017} \\
WASP~43~b & 0.93 & 0.0142 & 82.6 & 0.813475 & 0.6 & 4400 & 86.9633 & \cite{hellier2011} \\
WASP~121~b & 1.865 & 0.02544 & 87.6 & 1.2749255 & 1.458 & 6460 & 272.013 & \cite{delrez2016} \\
WASP~18~b & 1.158 & 0.02034 & 85.0 & 0.94145181 & 1.222 & 6400 & 123.925 & \cite{southworth2010} \\
WASP~79~b & 1.53 & 0.0519 & 86.1 & 3.662392 & 1.51 & 6600 & 248.447 & \cite{brown2017} \\
WASP~80~b & 0.999 & 0.0344 & 89.02 & 3.06785234 & 0.586 & 4143 & 49.8591 & \cite{triaud2015} \\
WASP~17~b & 1.991 & 0.0515 & 86.83 & 3.7354380 & 1.572 & 6650 & 410.408 & \cite{anderson2011} \\
WASP~39~b & 1.27 & 0.0486 & 87.83 & 4.055259 & 0.895 & 5400 & 215.295 & \cite{faedi2011} \\
WASP~69~b & 1.057 & 0.04527 & 86.71 & 3.8681382 & 0.813 & 4715 & 50.0323 & \cite{bonomo2017} \\
 &  &  &  &  &  &  &  & \cite{wallack2019} \\
HD~189733~b & 1.138 & 0.03099 & 85.58 & 2.218573 & 0.756 & 5040 & 19.7752 & \cite{torres2008} \\
HD~209458~b & 1.359 & 0.04707 & 86.71 & 3.524746 & 1.155 & 6065 & 48.3688 & \cite{torres2008} \\
HD~149026~b & 0.654 & 0.04313 & 90.0 & 2.87588874 & 1.368 & 6160 & 76.0306 & \cite{torres2008} \\
HAT-P-1 b & 1.242 & 0.0553 & 86.11 & 4.46543 & 1.135 & 5975 & 159.706 & \cite{torres2008} \\
HAT-P-26 b & 0.565 & 0.04357 & 88.09 & 4.2345023 & 0.788 & 5011 & 142.408 & \cite{wakeford2017} \\
 &  &  &  &  &  &  &  & \cite{wallack2019} \\
GJ~436~b & 0.3767 & 0.02872 & 86.36 & 2.6438983 & 0.464 & 3350 & 9.7559 & \cite{torres2008} \\
GJ~3470~b & 0.34615 & 0.031 & 88.88 & 3.3366487 & 0.48 & 3652 & 29.4463 & \cite{biddle2014} \\
 &  &  &  &  &  &  &  & \cite{benneke2019} \\
WASP~77A~b & 1.21 & 0.024 & 89.4 & 1.3600309 & 0.955 & 5500 & 105.465 & \cite{maxted2013} \\
WASP~52~b & 1.27 & 0.0272 & 85.35 & 1.7497798 & 0.79 & 5000 & 175.691 & \cite{hebrard2013} \\
WASP~127~b & 1.41 & 0.0522 & 88.1 & 4.178062 & 1.33 & 5750 & 160.233 & \cite{lam2017} \\
WASP~107~b & 0.94 & 0.055 & 89.7 & 5.721490 & 0.66 & 4430 & 64.8614 & \cite{anderson2017} \\
TOI~193~b & 0.45 & 0.0169 & 75.4 & 0.7921 & 0.99 & 5222 & 80.6231 & \cite{pearson2019} \\
GJ~1214~b & 0.254 & 0.01411 & 88.17 & 1.58040456 & 0.216 & 3026 & 14.6487 & \cite{harpsoe2013} \\
GJ~357~b & 1.217 & 0.035 & 89.12 & 3.93072 & 0.337 & 3505 & 9.4444 & \cite{luque2019} \\
GJ~1132~b & 1.16 & 0.0154 & 88.64 & 1.628930 & 0.207 & 3270 & 12.6176 & \cite{berta-thompson2015} \\
L98-59 c & 0.12023 & 0.0324 & 89.3 & 3.690621 & 0.312 & 3367 & 10.6226 & \cite{kostov2019} \\
L98-59 d & 0.14027 & 0.052 & 88.5 & 7.45086 & 0.312 & 3367 & 10.6226 & \cite{kostov2019} \\
LP~791-18~b & 0.09992 & 0.00969 & 87.3 & 0.9480050 & 0.171 & 2960 & 26.5128 & \cite{crossfield2019} \\
Trappist-1 b & 0.101914 & 0.011534 & 89.28 & 1.51088432 & 0.1234 & 2520 & 12.4299 & \cite{ducrot2020} \\
Trappist-1 d & 0.072085 & 0.02226 & 89.65 & 4.04978035 & 0.1234 & 2520 & 12.4299 & \cite{ducrot2020} \\
Trappist-1 e & 0.084263 & 0.02924 & 89.663 & 6.09956479 & 0.1234 & 2520 & 12.4299 & \cite{ducrot2020} \\
Trappist-1 f & 0.096128 & 0.038774 & 89.666 & 9.20659399 & 0.1234 & 2520 & 12.4299 & \cite{ducrot2020}
\enddata
\tablenotetext{a}{The distances of the systems have been taken from the Gaia Data Release 2 catalog \citep{gaiadr2}.}
\end{deluxetable*}

Most configurations for WASP~12~b have a positive spectrum bias; i.e., the phase-blend bias is larger than the self-blend bias. In this case, the largest bias occurs with the minimum circulation efficiency.
The dominant effect for WASP~43~b depends on the circulation efficiency. If $\varepsilon \sim 0.5$, the spectrum bias has both positive and negative values.
In general, the phase-blend component of the bias can be largely removed with conventional detrending practices, albeit by mixing astrophysical signals with instrumental systematic effects \citep{martin-lagarde2020}. It is worth pointing out that the correction of the phase-blend effect must be wavelength-dependent, as a white light curve correction would not attenuate the spectral bias. The remaining self-blend bias increases monotonically in absolute value with the atmospheric circulation efficiency. The self-blend effect can only be corrected with full phase curve observations or by relying on theoretical models.

Figure \ref{fig:total_bias_bb_ace} compares the spectroscopic biases obtained with blackbody and equilibrium chemistry spectra for the exoplanet dayside and nightside. These latter spectra can introduce some molecular features in addition to the smooth trend obtained with blackbody spectra.

\begin{deluxetable*}{ccccccc}[t!]
\tablecaption{Adopted $T_{\text{day}}$, $T_{\text{night}}$, $A_b$, and $\varepsilon$ based on the available observations, and/or for the extreme cases that maximise the phase- and self-blend, respectively. \label{tab:T_Ab_eps}}
\tablecolumns{7}
%\tablenum{1}
\tablewidth{0pt}
\tablehead{
 & \multicolumn4c{Most likely configurations} & \multicolumn2c{Extreme configurations} \\
\colhead{Planet} & \colhead{Obs. Type}\tablenotemark{a} & \colhead{$T_{\text{day}}$/$T_{\text{night}}$ ($K$)} & \colhead{$A_b$, $\varepsilon$} & \colhead{References} & \colhead{$T_{\text{day}}$ ($T_{\text{night}}=0 \, K$)} & \colhead{$T_{\text{day}}=T_{\text{night}}$ ($K$)} \\
 & & & & & $A_b=0$, $\varepsilon=0$ & $A_b=0$, $\varepsilon=1$
}
\startdata
WASP~12~b & phase curve & 2850/1350 & 0.40, 0.12 & \cite{bell2019} & 3299 & 2581 \\
WASP~43~b & phase curve & 1600/850 & 0.23, 0.19 & \cite{morello2019} & 1762 & 1379 \\
WASP~121~b & phase curve & 2950/1170 & 0.044, 0.064 & \cite{daylan2019} & 3014 & 2358 \\
WASP~18~b & phase curve & 2890/480 & 0.20, 0.002 & \cite{arcangeli2019} & 3057 & 2392 \\
WASP~79~b & eclipse & 1950/1510 & 0.0, 0.60 & \cite{garhart2020} & 2194 & 1717 \\
WASP~80~b & eclipse & 890/770 & 0.0, 0.77 & \cite{triaud2015} & 1054 & 825 \\
WASP~17~b & eclipse & 1730/1730 & 0.091, 1.0 & \cite{anderson2011} & 2264 & 1772 \\
WASP~39~b & eclipse & 1130/1110 & 0.0, 0.97 & \cite{kammer2015} & 1428 & 1117 \\
WASP~69~b & eclipse & 940/940 & 0.094, 1.0 & \cite{wallack2019} & 1231 & 964 \\
HD~189733~b & phase curve & 1250/990 & 0.27, 0.63 & \cite{knutson2009,knutson2012} & 1534 & 1200 \\
HD~209458~b & phase curve & 1500/970 & 0.44, 0.36 & \cite{zellem2014} & 1851 & 1449 \\
HD~149026~b & phase curve & 1800/1030 & 0.41, 0.24 & \cite{zhang2018} & 2138 & 1673 \\
HAT-P-1 b & eclipse & 1500/1120 & 0.0, 0.55 & \cite{todorov2010} & 1668 & 1305 \\
HAT-P-26 b & eclipse & 700/700 & 0.78, 1.0 & \cite{wallack2019} & 1313 & 1028 \\
GJ~436~b & eclipse & 790/480 & 0.0, 0.28 & \cite{stevenson2010} & 830 & 649 \\
GJ~3470~b & eclipse & 510/510 & 0.71, 1.0 & \cite{benneke2019} & 886 & 693 \\
WASP~77A~b & none & -- & -- & -- & 2138 & 1673 \\
WASP~52~b & none & -- & -- & -- & 1660 & 1299 \\
WASP~127~b & none & -- & -- & -- & 1788 & 1400 \\
WASP~107~b & none & -- & -- & -- & 946 & 740 \\
TOI~193~b & none & -- & -- & -- & 2463 & 1927 \\
GJ~1214~b & none & -- & -- & -- & 730 & 571 \\
GJ~357~b & none & -- & -- & -- & 670 & 524 \\
GJ~1132~b & none & -- & -- & -- & 739 & 578 \\
L98-59 c & none & -- & -- & -- & 644 & 504 \\
L98-59 d & none & -- & -- & -- & 508 & 398 \\
LP~791-18 d & none & -- & -- & -- & 766 & 600 \\
Trappist-1 b & none & -- & -- & -- & 508 & 397 \\
Trappist-1 d & none & -- & -- & -- & 366 & 286 \\
Trappist-1 e & none & -- & -- & -- & 319 & 250 \\
Trappist-1 f & none & -- & -- & -- & 277 & 217 \\
\hline
\enddata
\tablenotetext{a}{Type of observations used to estimate $T_{\text{day}}$ and $T_{\text{night}}$ (most likely configurations). Whereas multiple observations were available, we took the arithmetic average of the brightness temperatures reported in the reference paper. The eclipse observations provide $T_{\text{day}}$ only; the corresponding $T_{\text{night}}$ was calculated by assuming the maximum circulation efficiency compatible with the eclipse data.}
\end{deluxetable*}

\section{Biased retrieval analyses} \label{sec:biased_retrievals}

In this section, we describe the simulations aimed at characterizing the scientific impact of potential bias in the transmission spectra of exoplanet atmospheres observed with JWST.
The essential steps of this study are:
\begin{enumerate}
\item creating the reference transmission spectra,
\item computing the spectroscopic bias and the corresponding biased spectra under different hypotheses,
\item performing atmospheric retrievals on the biased transmission spectra, and
\item comparing the retrieved atmospheric properties with those of the reference transmission spectra.
\end{enumerate}

We selected two targets among those that showed potentially significant bias in infrared transit photometry according to \cite{martin-lagarde2020}.
\begin{itemize}
\item WASP~12~b has the largest predicted bias (positive) in their sample. It is the subject of several theoretical studies and past observations \citep{cowan2012,stevenson2014b,stevenson2014c,kreidberg2015,bell2019}.
\item WASP~43~b can have one of the largest positive or negative biases, depending on its atmospheric circulation efficiency. It is the subject of many theoretical studies and past and future observations \citep{stevenson2014,stevenson2017,keating2017,bean2018,mendonca2018, morello2019,may2020,venot2020}.
\end{itemize}
These planets represent hot Jupiters with ultrashort orbital periods around different types of host stars.% (see Table \ref{tab:system_parameters}).

\subsection{Reference transmission spectra}

We took the transmission spectra observed with HST/WFC3 as starting points. In particular, we reanalyzed the low-resolution spectra at 1.1-1.7 $\mu$m reported by \cite{tsiaras2018} using the novel Tau-REx III retrieval algorithm \citep{al-refaie2019}. 
As the narrow wavelength range limits the ability to detect molecules other than H$_2$O, we adopted atmospheric chemical equilibrium (ACE) models \citep{agundez2012, venot2012} instead of the most popular free chemistry to include more absorbing species. The molecular absorption cross sections were computed by the ExoMol group \citep{chubb2020}. We also included collisionally induced absorption (CIA) by H$_2$--H$_2$ and H$_2$--He \citep{gordon2017} and Mie scattering \citep{lee2013}. We assumed an isothermal temperature profile.

The free parameters in our retrievals were the planet radius at 10 bar, the atmospheric temperature, Mie scattering cloud particle size, mixing ratio, and composition. The carbon-to-oxygen (C/O) ratio and metallicity were fixed to solar values. The best-fit solutions obtained with these settings have the same Bayesian evidence as those reported by \cite{tsiaras2018} to within $\Delta \log{E} \le 2$, despite using only five free parameters instead of 12 and 10 for WASP~12~b and WASP~43~b, respectively.

We calculated forward model spectra that extend the best-fit solutions over a broader wavelength range, then bin-averaged. In this way, we obtained the transmission spectra associated with three JWST instruments using the same observing modes planned for the Transiting Exoplanet Community ERS program \citep{bean2018}:
\begin{itemize}
\item the Near-InfraRed Imager and Slitless Spectrograph (NIRISS) with the Single-Object Slitless Spectroscopy (SOSS) mode,
\item the Near-InfraRed Spectrograph (NIRSpec) with the G235H and G395H gratings, and
\item the MIRI with the Low-Resolution Spectroscopy (LRS) slitless mode.
\end{itemize}
Table \ref{tab:jwst_obs_modes} reports the nominal wavelength ranges for the three instruments and the size of the spectroscopic bins. 
Figure \ref{fig:PCEs} shows the corresponding PCEs. The PCE for each observing setup is the combined throughput from the telescope, instrument optics, detector efficiency, quantum efficiency, and filter throughput (if applicable). We adopted the reference data files from the Pandeia Exposure Time Calculator \citep{pontoppidan2016}, and precalculated response functions for the MIRI LRS slitless \citep{kendrew2015} and NIRISS SOSS \citep{maszkiewicz2017}. The resulting PCEs are among the passbands available within the \texttt{ExoTETHyS} package.

Additionally, we considered another set of reference transmission spectra with identical parameters, except for  C/O=1.5.

\begin{figure*}[t!]
\includegraphics[width=\textwidth]{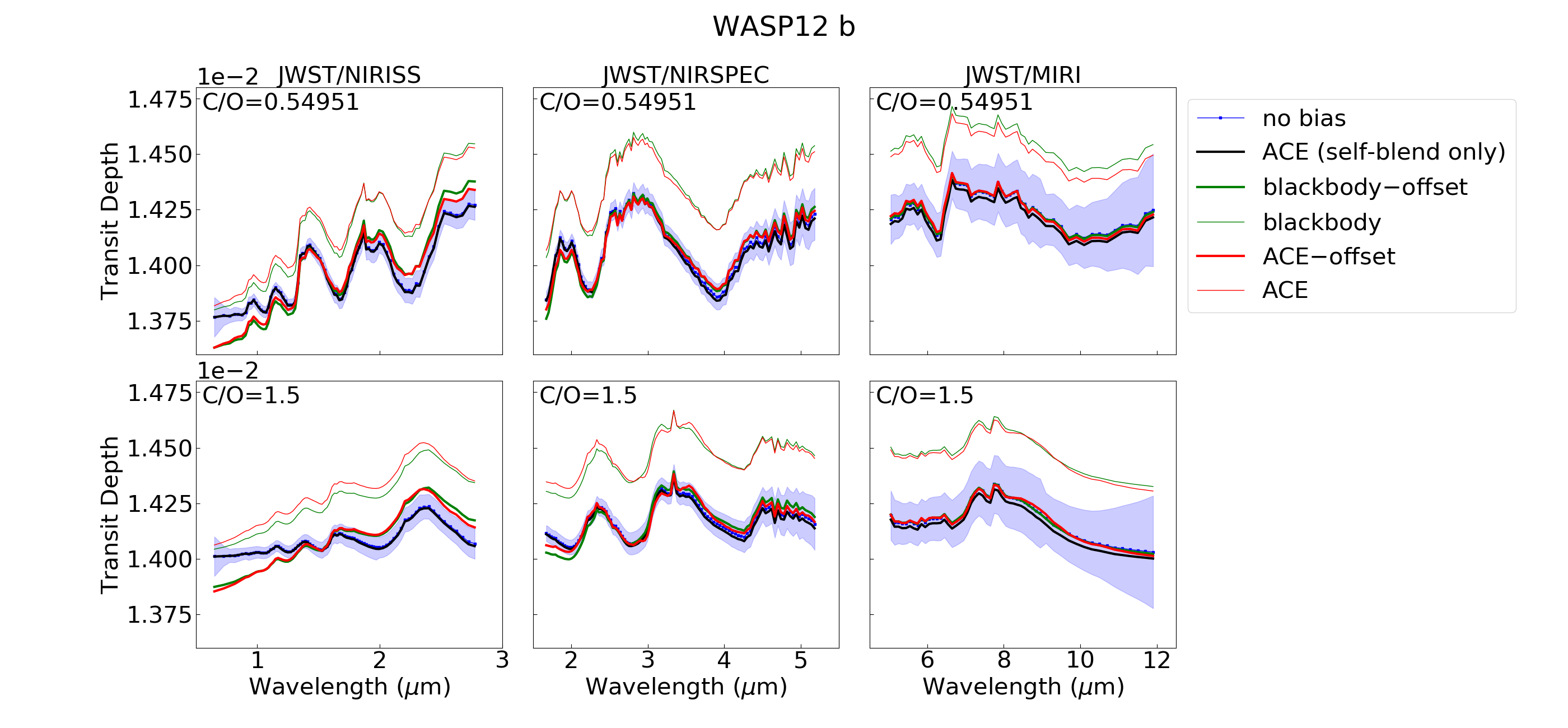}
\includegraphics[width=\textwidth]{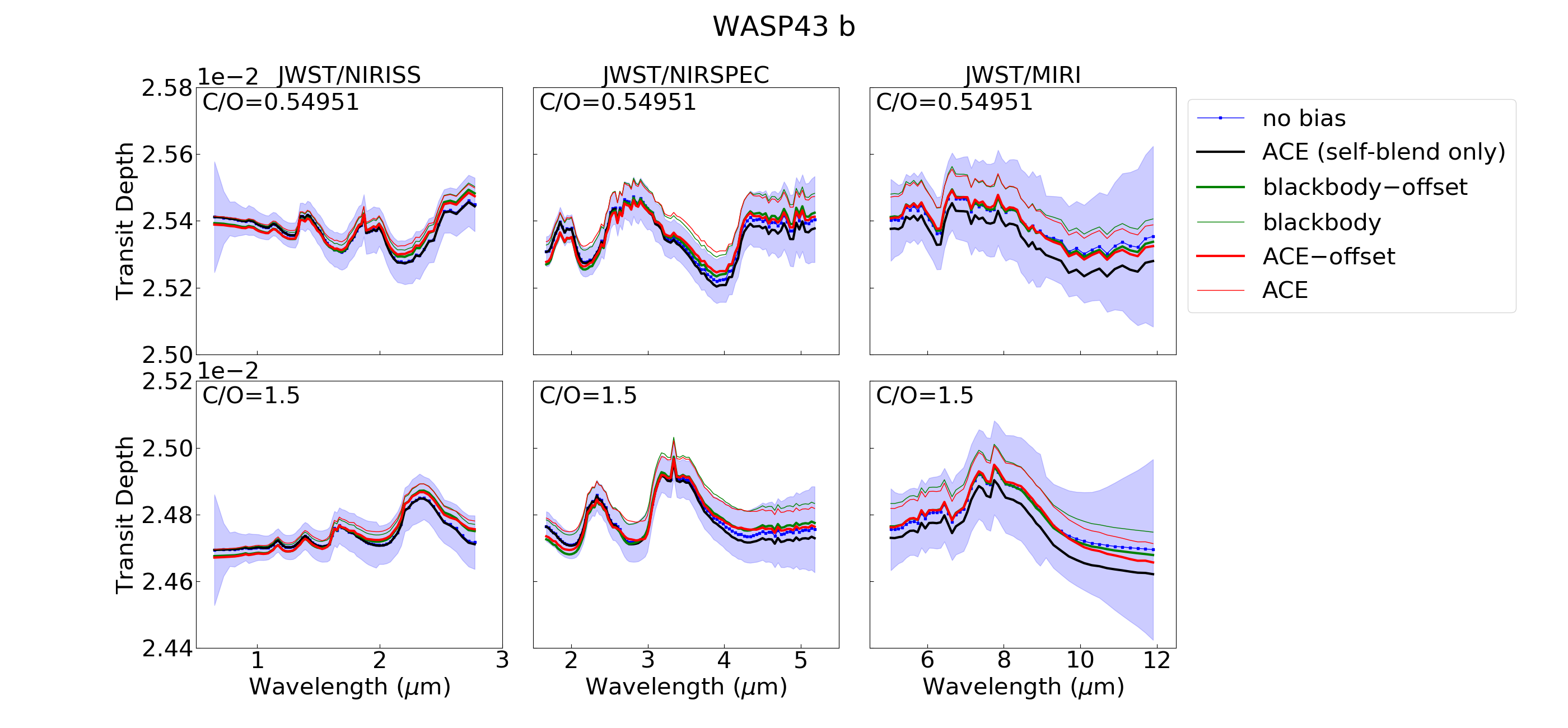}
\caption{Simulated transmission spectra for WASP~12~b and WASP~43~b. Each panel contains a set of spectra for a specific JWST instrument (from left to right: NIRISS, NIRSpec, and MIRI) and atmospheric C/O ratio (upper rows: solar C/O; lower rows: C/O=1.5). The pure transmission spectra (no bias added) are in blue, with a swath denoting the 1$\sigma$ error bars estimated by using \texttt{ExoTETHyS.BOATS}. The other spectra are biased as described in Section \ref{ssec:spetrum_bias}. The ``model$-$offset'' assumes a constant correction that does not account for the wavelength dependence of the bias. A significantly biased spectral slope is observed for WASP~12~b with NIRISS ($\gg$1$\sigma$), and in the bluer part of the NIRSPEC spectra $\gtrsim$1$\sigma$), and there is a large but nearly constant offset with MIRI. The biased spectra for WASP~43~b are almost entirely within the 1$\sigma$ error bars. \label{fig:biased_spectra}}
\end{figure*}

\begin{figure*}[t!]
\includegraphics[width=0.8\textwidth]{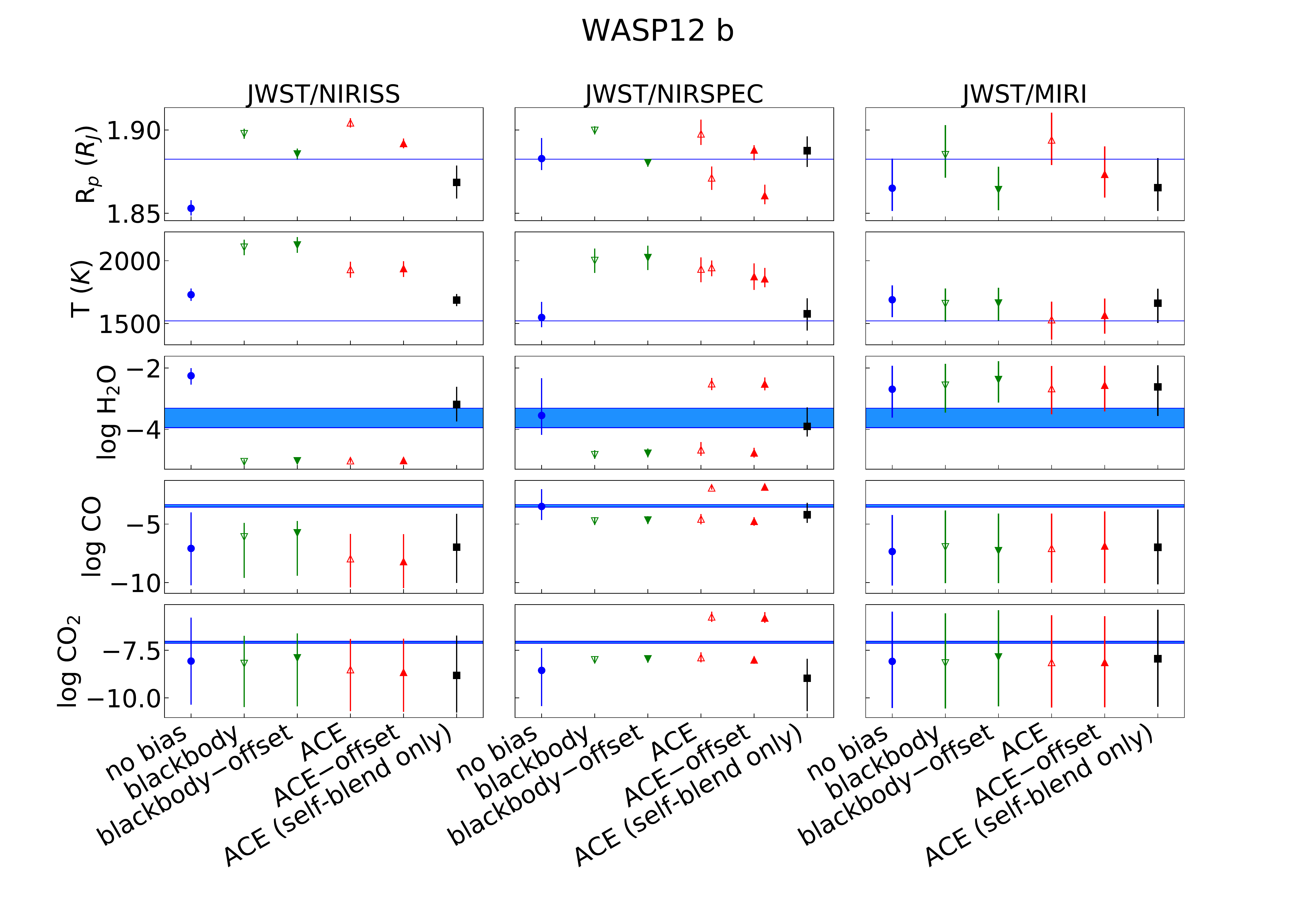}
\includegraphics[width=0.8\textwidth]{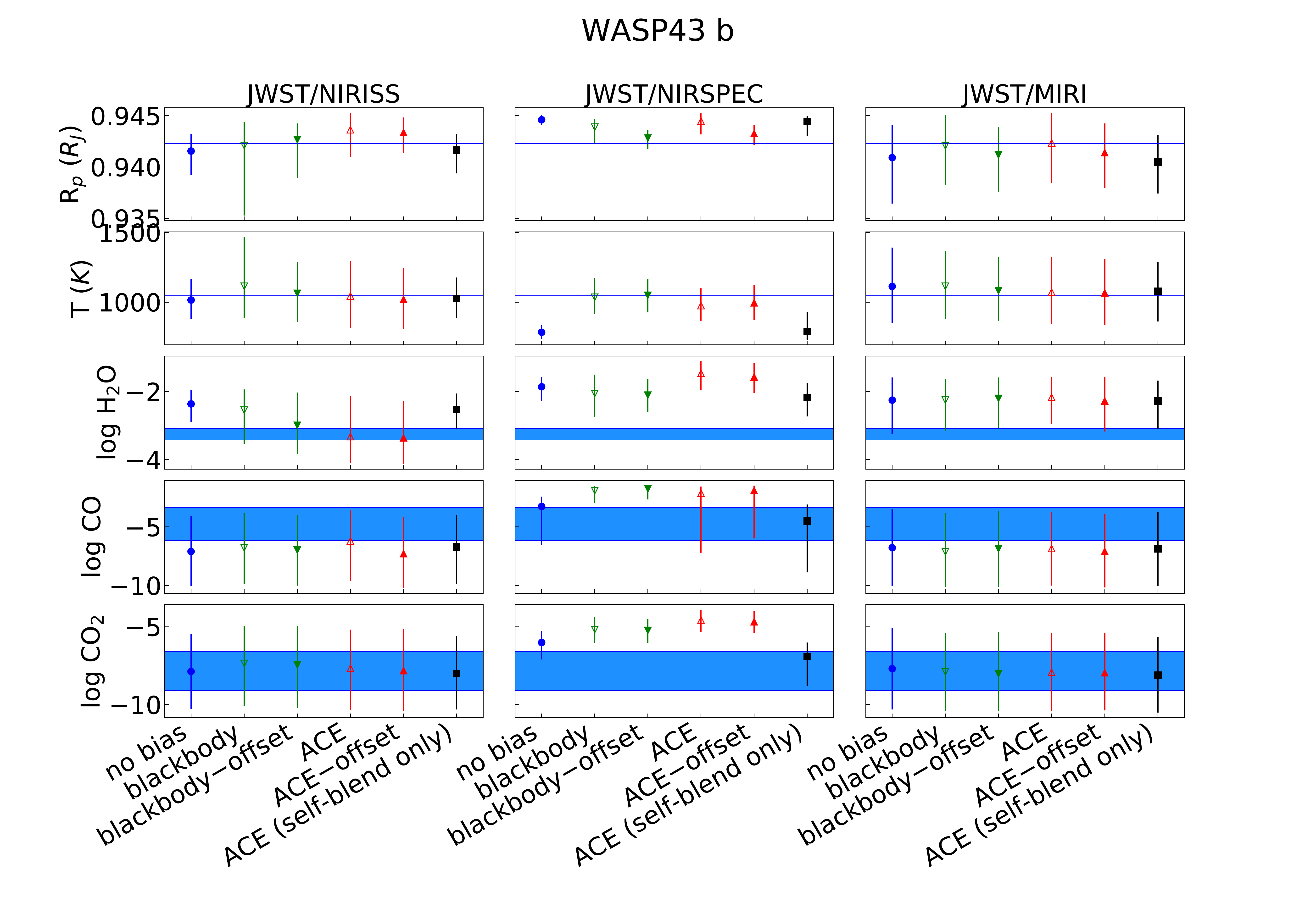}
\caption{Parameters retrieved from the simulated transmission spectra for WASP~12~b and WASP~43~b shown in Figure \ref{fig:biased_spectra} (cases with with solar C/O ratio). For the NIRSpec spectra of WASP~12~b with ACE and ACE$-$offset bias, the retrieval algorithm found two statistically equivalent solutions, which are both reported. The blue horizontal lines indicate the input parameter values, and the shaded areas indicate the range of molecular mixing ratios at various heights in the atmosphere (the mixing ratios are kept constant in the retrievals). Where the spectral slope was largely biased, the retrieved atmospheric temperature and water mixing ratio can be significantly different from those obtained from the unbiased spectra (WASP~12~b; NIRISS and NIRSpec). A constant offset appears to only affect the inferred planet's radius (WASP~12~b; MIRI). In a few cases, the parameters retrieved from the unbiased spectra do not match the input values, likely due to approximations adopted in the retrievals or degeneracies. \label{fig:biased_retrievals}}
\end{figure*}

\subsection{Biased transmission spectra}
\label{ssec:spetrum_bias}

We used \texttt{ExoTETHyS.BOATS} to calculate the bias in transit depth on the same wavelength bins as the transmission spectra. Table \ref{tab:system_parameters} reports the system parameters. Table \ref{tab:T_Ab_eps} contains the dayside and nightside temperatures based on current estimates from observations. We modeled the emission from the two planetary hemispheres in the following ways:
\begin{itemize}
\item assuming blackbody spectra with $T_{\text{day}}$ and $T_{\text{night}}$ from Table \ref{tab:T_Ab_eps} (most likely configurations) and
\item calculating the emission spectra with Tau-REx III, assuming ACE with solar C/O ratio (and C/O=1.5) and metallicity, CIA, Rayleigh scattering, and temperature profiles as in \cite{guillot2010} with the same $T_{\text{day}}$ and $T_{\text{night}}$.
\end{itemize}
We then added the spectroscopic bias in three forms to the pure transmission spectra to obtain the biased spectra:
\begin{itemize}
\item total bias resulting from the self- and phase-blend effects;
\item total bias minus average offset, such that the sum of the biases on the wavelength bins is zero; and
\item self-blend bias only.
\end{itemize}
The case of total bias minus average offset may occur {if a constant correction is applied at all wavelengths. For example, when the spectral light curves are divided by the white one (e.g., \citealp{tsiaras2016b}), if the bias in transit depth is removed for white light curve, the spectral values will also be shifted, but they remain biased. The self-blend bias is what remains if the phase-blend effect is removed from the data for each wavelength \citep{martin-lagarde2020}.
Figure \ref{fig:biased_spectra} shows the pure transmission and biased spectra considered for WASP~12~b and WASP~43~b.
The biased spectra appear to have an offset and a different slope than the pure transmission ones in the NIRISS passband. The effect on the slope is smaller at the longer wavelengths, and it is negligible in the MIRI passband. The bias for WASP~12~b is much larger than the 1$\sigma$ error bars in most configurations, while for WASP~43~b it is within about 1$\sigma$.

\subsection{Atmospheric retrievals}

We performed atmospheric retrievals on the pure and biased transmission spectra using Tau-REx III. The free fitting parameters were the planet radius, the isothermal temperature, three Mie scattering parameters, and the abundances of all of the molecules used by the ACE module to compute the transmission spectra. We take the approximation of uniform chemical distribution with height for the retrievals, neglecting the pressure dependence of the input models.

\subsection{Biases in retrieval results}

In all cases, the retrieval algorithm found a solution that provides an excellent match to the simulated spectrum, albeit with different than input parameters.
Figure \ref{fig:biased_retrievals} compares the atmospheric parameters retrieved from pure and biased transmission spectra with their underlying values and/or intervals from the input models.
The spectrum bias can affect the H$_2$O abundance measured on WASP~12~b by orders of magnitude, the temperature by $\sim$500 $K$, and the planet radius by 2-3$\%$. In some cases, the measured discrepancies may include the effect of parameter degeneracies, which are discussed below. The largest effects occur with JWST/NIRISS, for which the biased spectra have a significantly different slope than the pure transmission ones. We obtained similar trends using blackbody or ACE spectra for planetary emission, but it is possible to have larger differences in the presence of more pronounced molecular features. The constant offsets change only the apparent radius of the planet, but not the other atmospheric parameters.

Unsurprisingly, the parameters obtained for WASP~43~b are not significantly affected by the modest spectroscopic bias. If the phase-blend effect is corrected, the self-blend bias is negligible for both WASP~12~b and WASP~43~b.

We note that, in some cases, the retrieved parameters do not match the underlying values even in the absence of spectrum bias. These discrepancies can only be partially explained by the approximation of constant gas profiles adopted in the retrievals \citep{changeat2019}. The error bars obtained from atmospheric retrievals can sometimes be underestimated, not accounting for the full parameter degeneracies \citep{rocchetto2016}. Performing retrievals over the broader wavelength range covered by the three instruments should help to significantly reduce the degeneracies between the planet's radius and atmospheric parameters \citep{chapman2017,welbanks2019}. A discussion of the robustness of the retrievals themselves is beyond the scope of this paper.

\section{Predicted impact on future observations} \label{sec:jwst_gto_ers}

We calculated the bias spectra for all exoplanets that will be observed in the JWST GTO and ERS programs as described in Section \ref{ssec:spetrum_bias} with blackbody emission spectra. Table \ref{tab:system_parameters} reports the system parameters. Table \ref{tab:T_Ab_eps} reports the dayside and nightside temperatures based on current estimates from observations and their theoretical limits. Tables \ref{tab:spectral_biases} and \ref{tab:spectral_biases_extreme} show the mean offsets and peak-to-peak amplitudes of the spectroscopic bias for the different configurations.
For most planets, the spectroscopic bias is smaller than the 1$\sigma$ error bars. The only JWST GTO and ERS targets showing a potential bias larger than 3$\sigma$ with a single visit are WASP~121~b, WASP~18~b, WASP~77A~b, WASP~43~b, and HD189733~b (see Table \ref{tab:spectral_biases_extreme}). 
It is worth noting that phase-blend is the dominant bias component for four out of five planets, which can be largely removed by appropriate data detrending. Furthermore, the full phase curves of WASP121~b and WASP~43~b will be observed. The only target with a potentially significant self-blend bias is HD189733~b, but it should not affect the secondary eclipse observations that are currently scheduled.

Figures \ref{fig:colormap_Gstar_b0.5} and \ref{fig:colormap_Mstar_b0.5} provide color maps to highlight regions of the parameter space from which the exoplanet spectra observed with JWST are more likely to be affected by the self- and phase-blend effects. In general, we can have spectral biases larger than 10 ppm for Neptune-sized or larger planets. The phase-blend bias can be up to hundreds of ppm for hot Jupiters with periods shorter than 2 days. The self-blend bias is more weakly dependent on the orbital period, but it rarely exceeds the 100 ppm.

Future transmission spectra taken with the Ariel mission can also be affected by the dark planet approximation. The general trends inferred for the JWST spectra can be extended to the Ariel ones, with some quantitative differences due to the specific instruments. Ariel will simultaneously cover the spectral region 0.5-7.8 $\mu$m, thus comprising the wavelength range of JWST/NIRSpec, and partly also that of NIRISS and MIRI. The Ariel spectra will have lower resolution, especially at wavelengths shorter than 2 $\mu$m. Due to the different telescope apertures, the Ariel mission will require multiple visits to achieve the same signal-to-noise ratio that would be obtained with a single JWST visit.

\begin{figure*}[t!]
\includegraphics[width=0.49\textwidth]{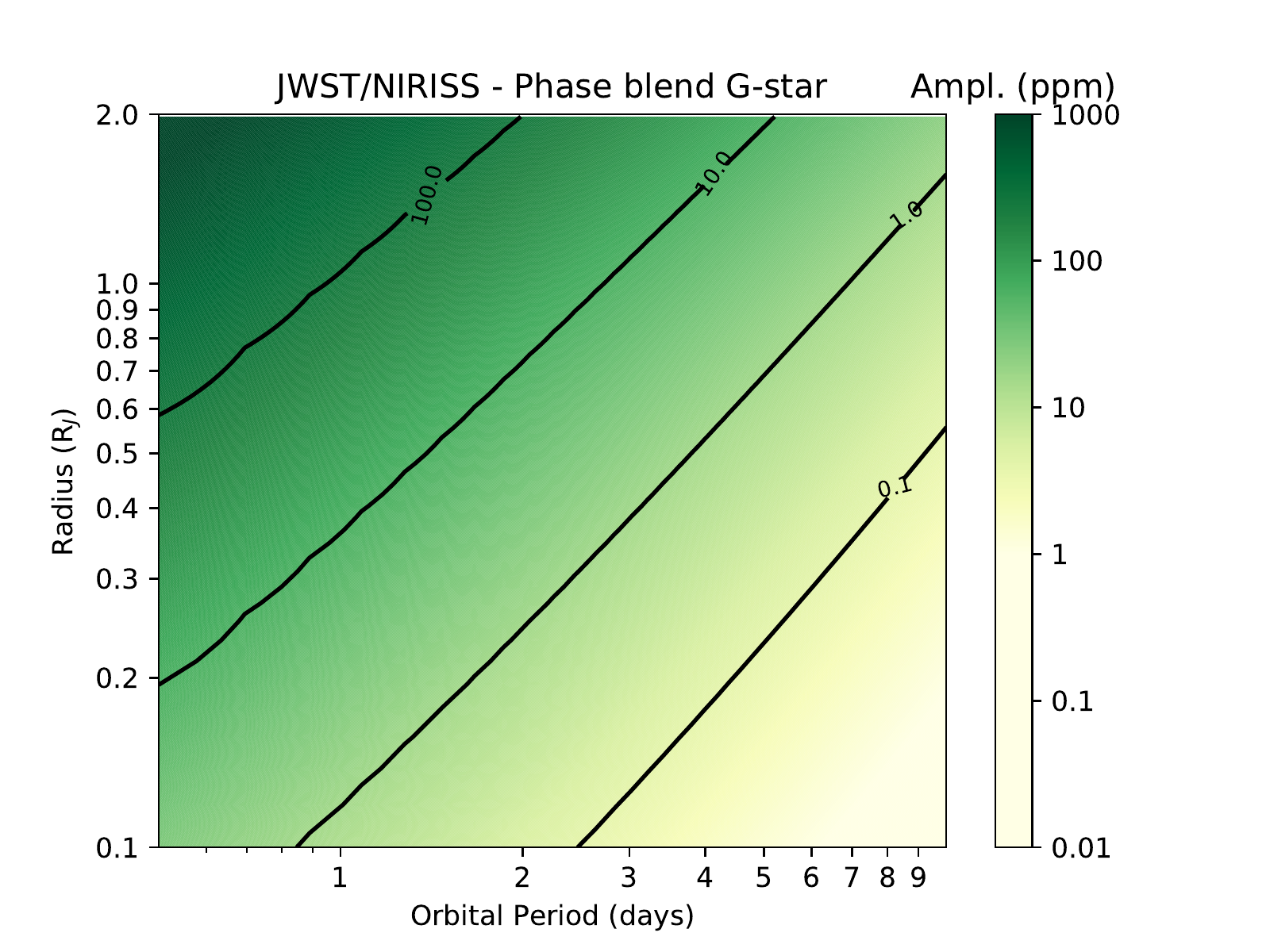}
\includegraphics[width=0.49\textwidth]{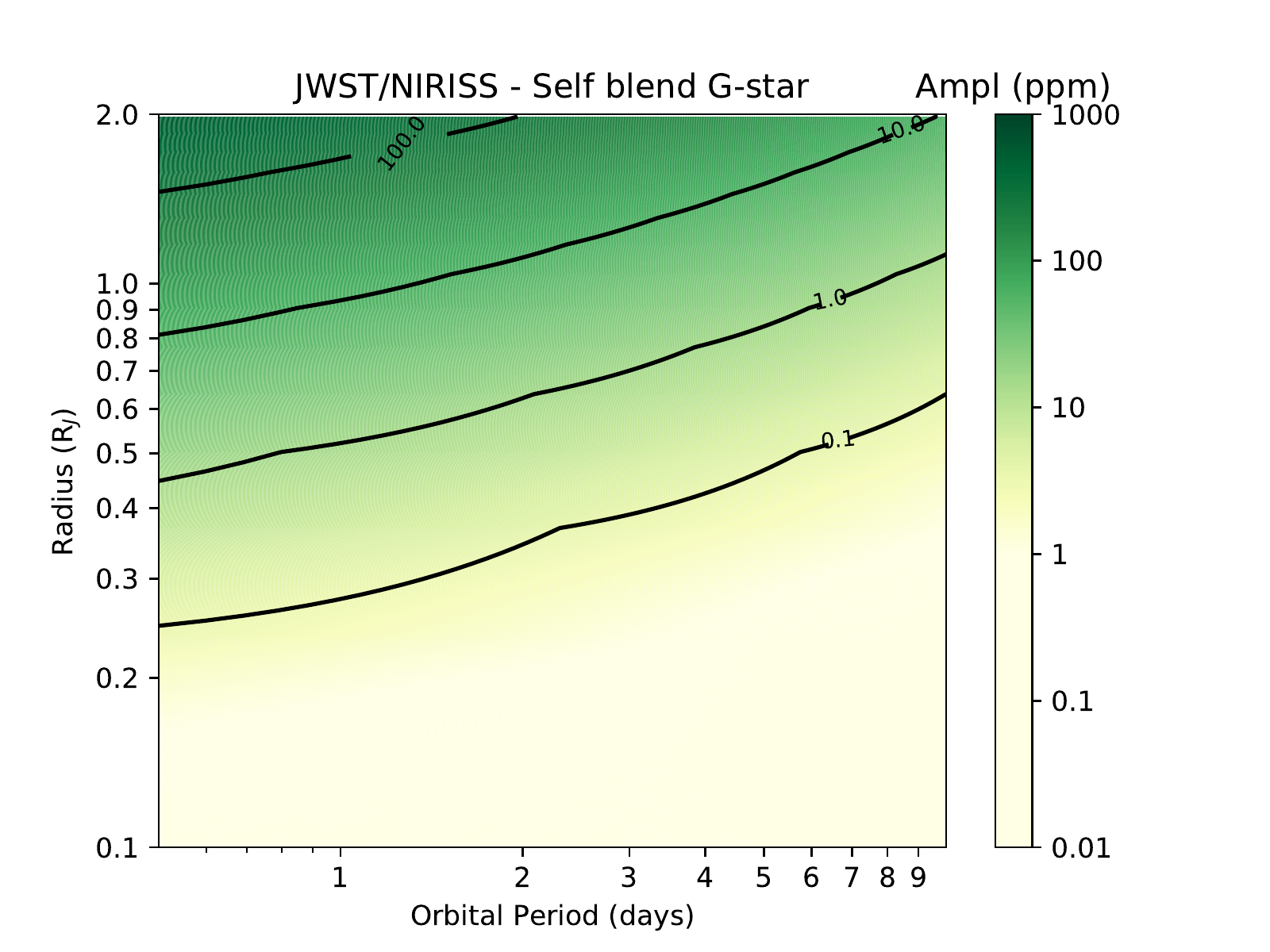}
\includegraphics[width=0.49\textwidth]{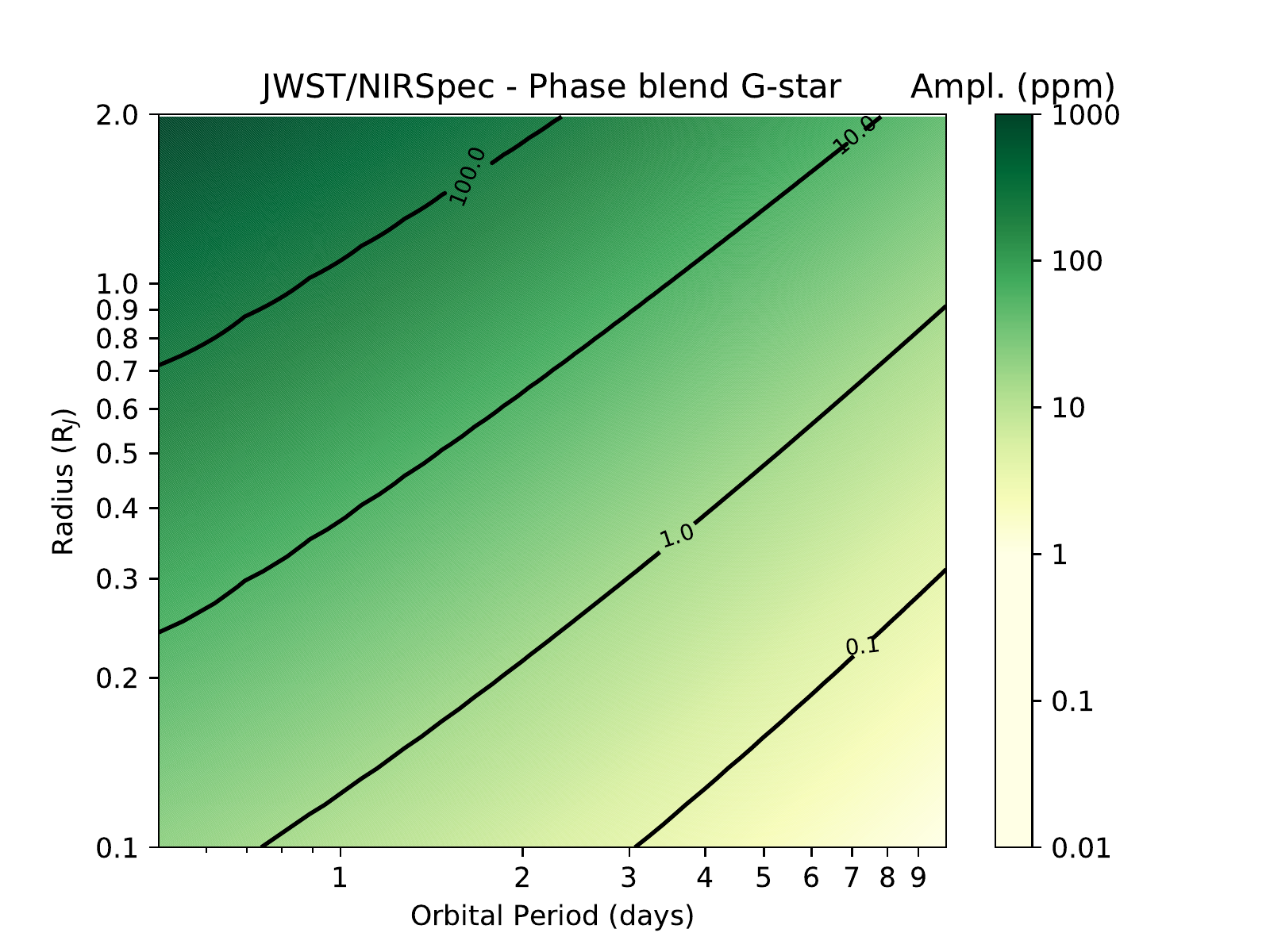}
\includegraphics[width=0.49\textwidth]{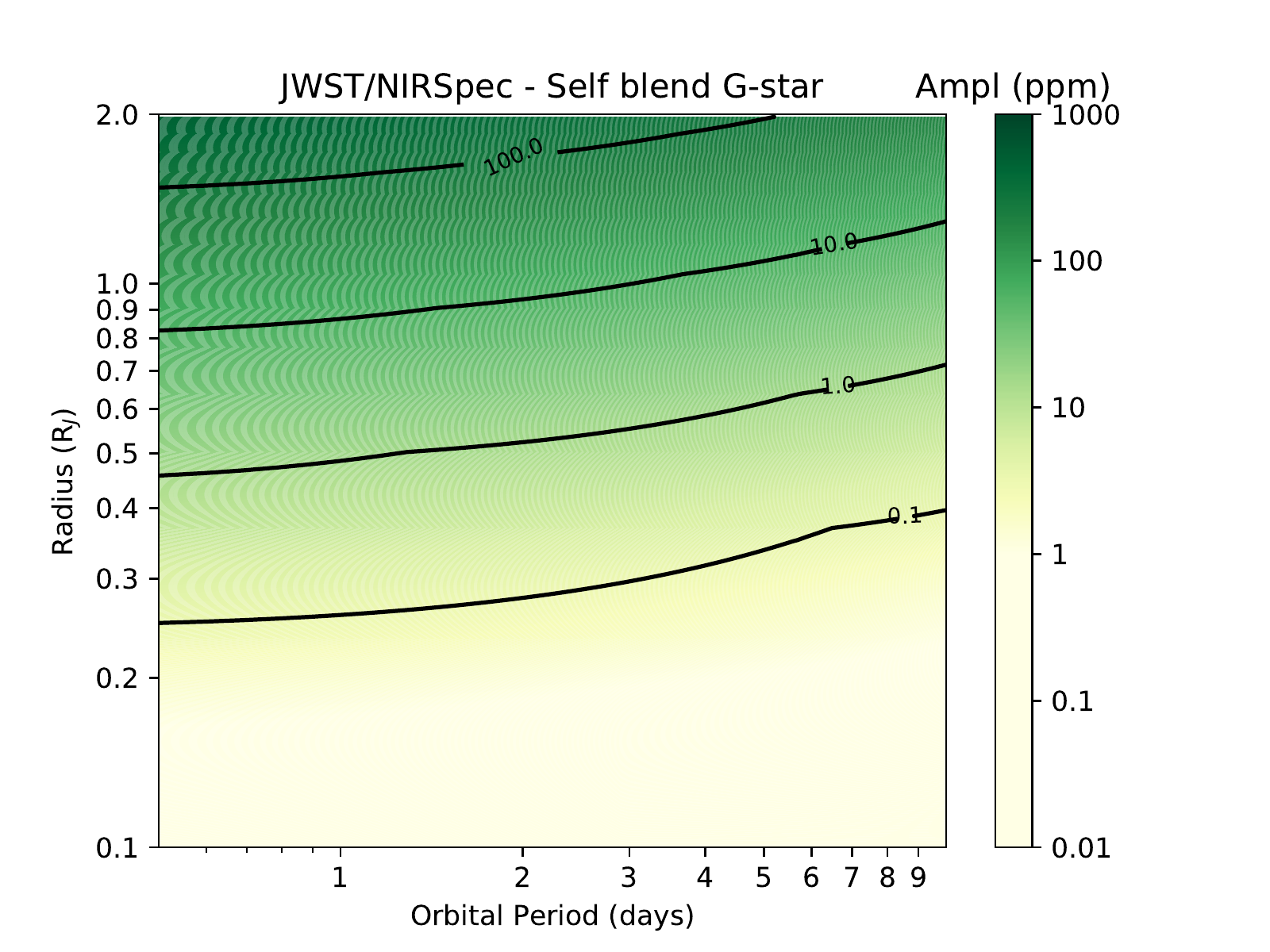}
\includegraphics[width=0.49\textwidth]{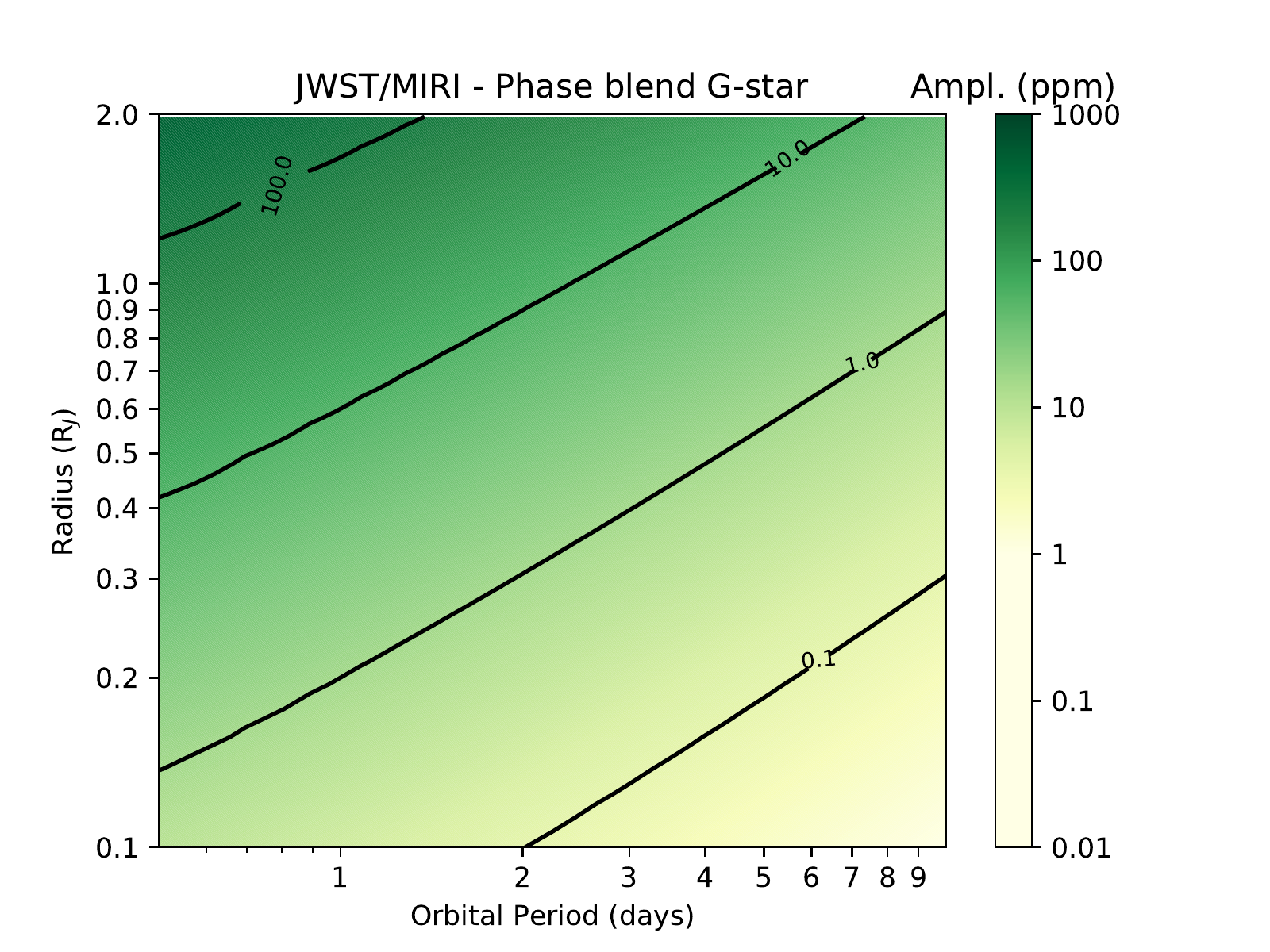}
\includegraphics[width=0.49\textwidth]{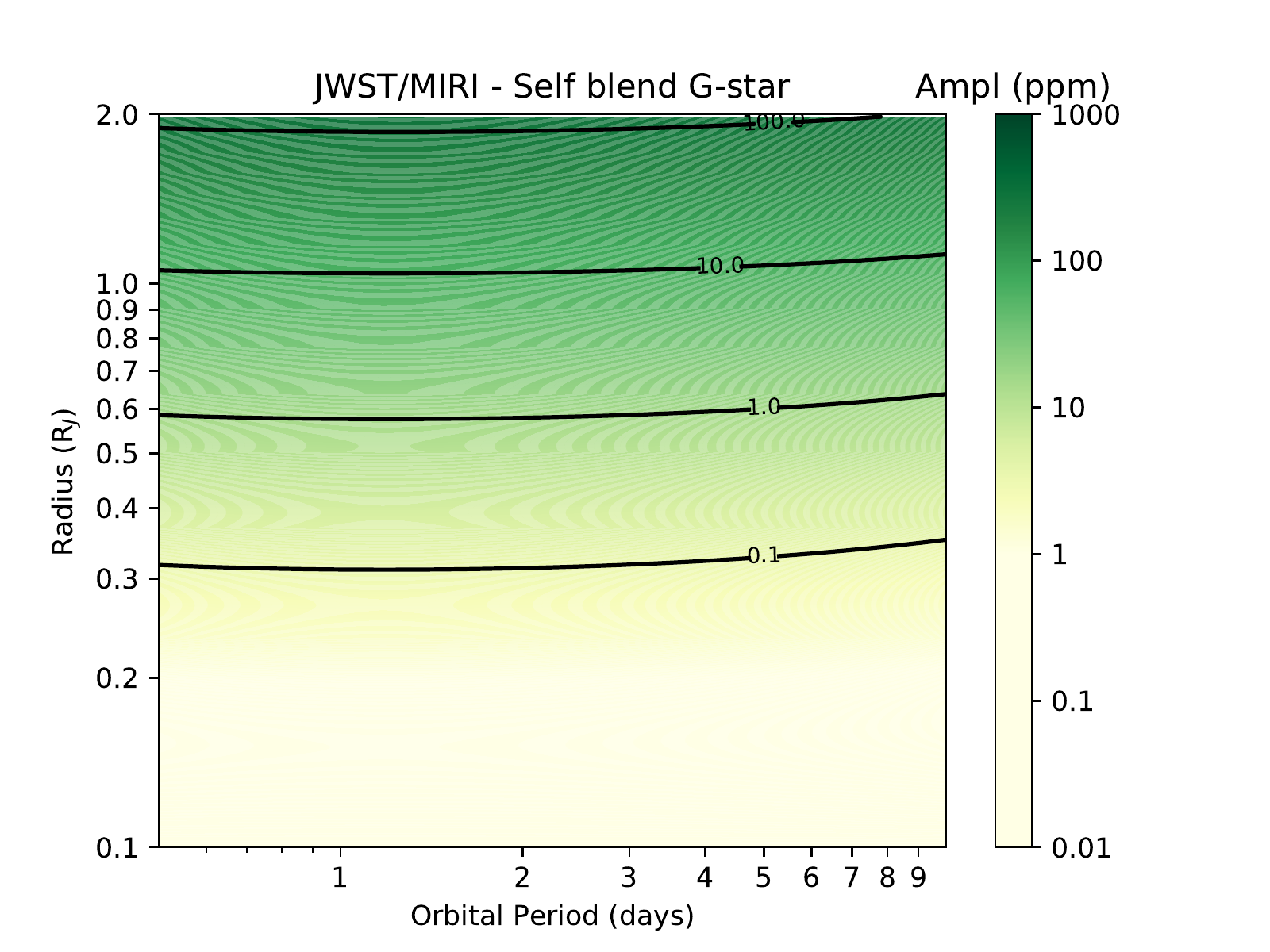}
\caption{Color maps with orbital period vs. planet's radius showing the peak-to-peak amplitude of the spectral biases for the JWST transmission spectra of exoplanets around a star with $T_{*,\text{eff}} = 5800 \, K$, solar mass and radius. We consider transits with impact parameter $b=0.5$, for homogeneity. The left column reports the results for the phase-blend bias (with $\varepsilon = 0.0$), and the right column reports those for the self-blend bias (with $\varepsilon = 1.0$).\label{fig:colormap_Gstar_b0.5}}
\end{figure*}

\begin{figure*}[t!]
\includegraphics[width=0.49\textwidth]{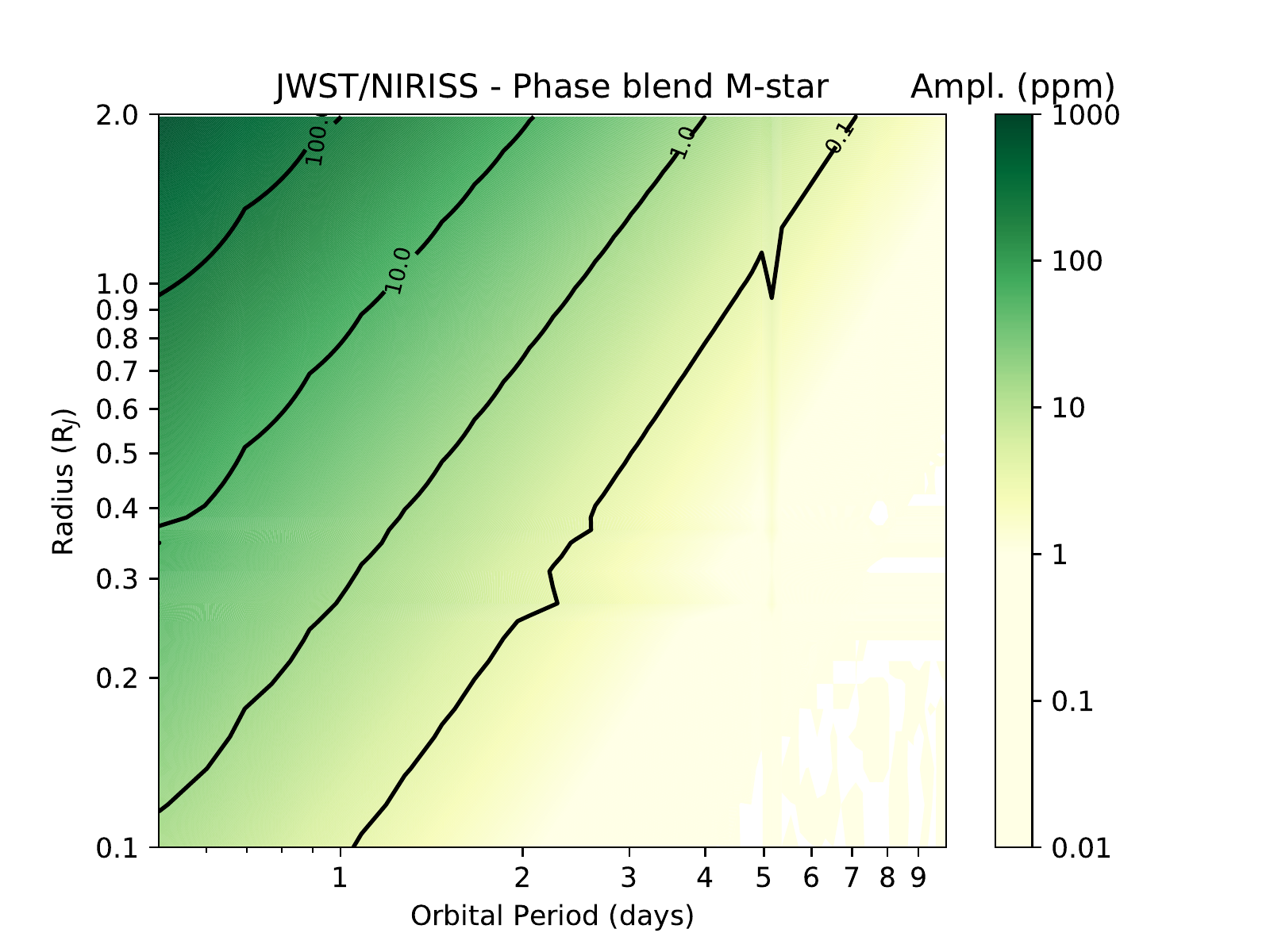}
\includegraphics[width=0.49\textwidth]{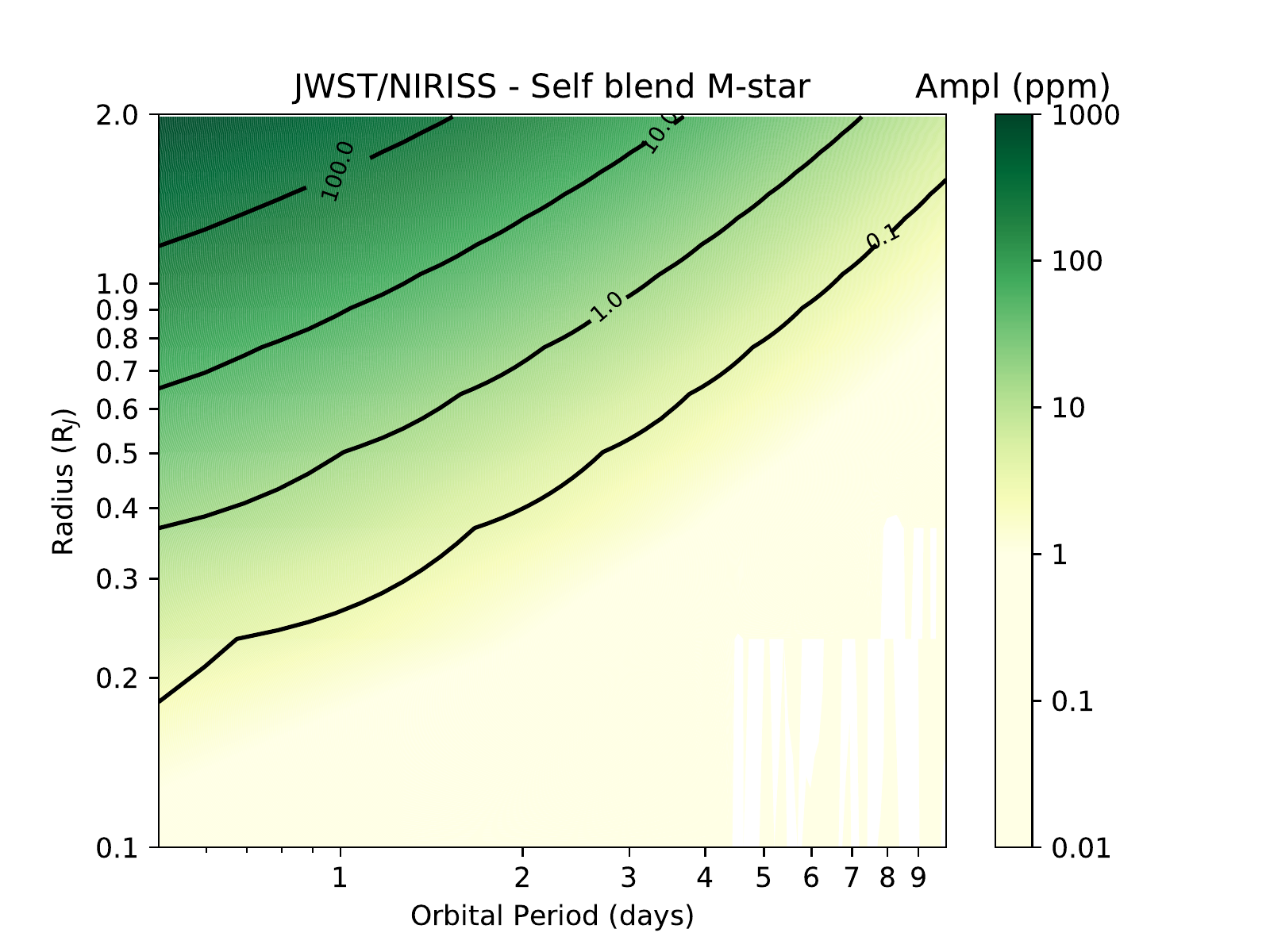}
\includegraphics[width=0.49\textwidth]{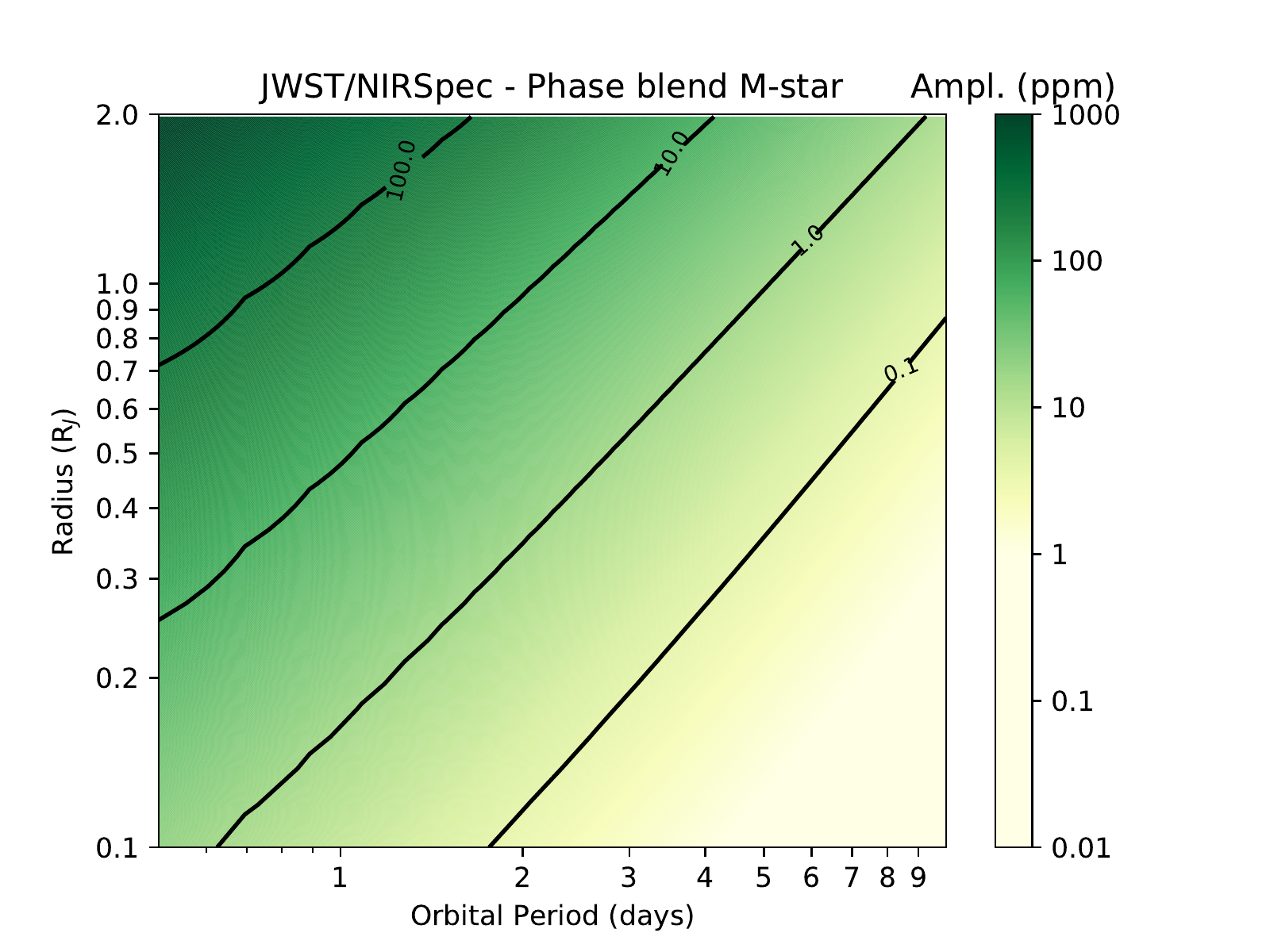}
\includegraphics[width=0.49\textwidth]{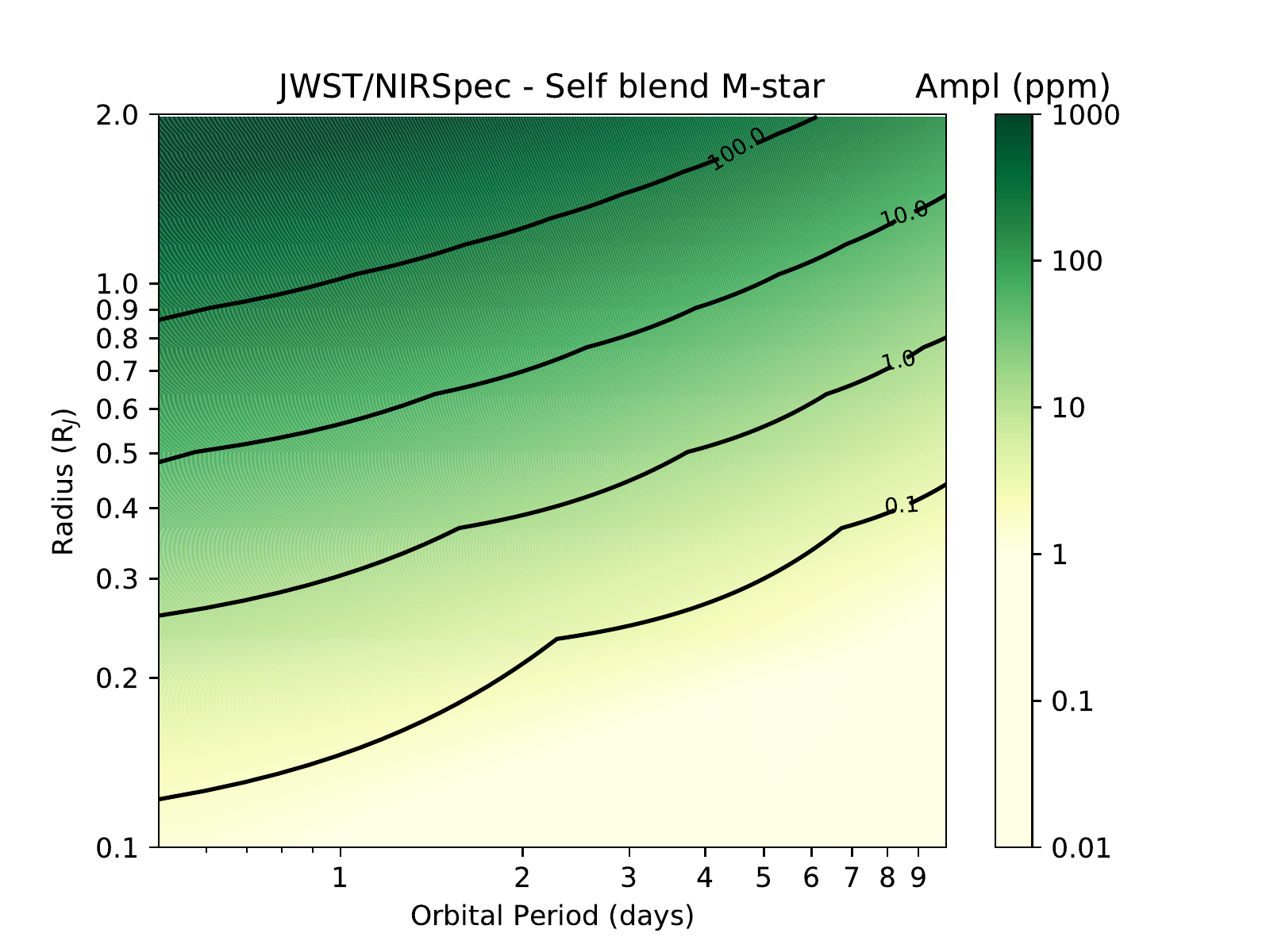}
\includegraphics[width=0.49\textwidth]{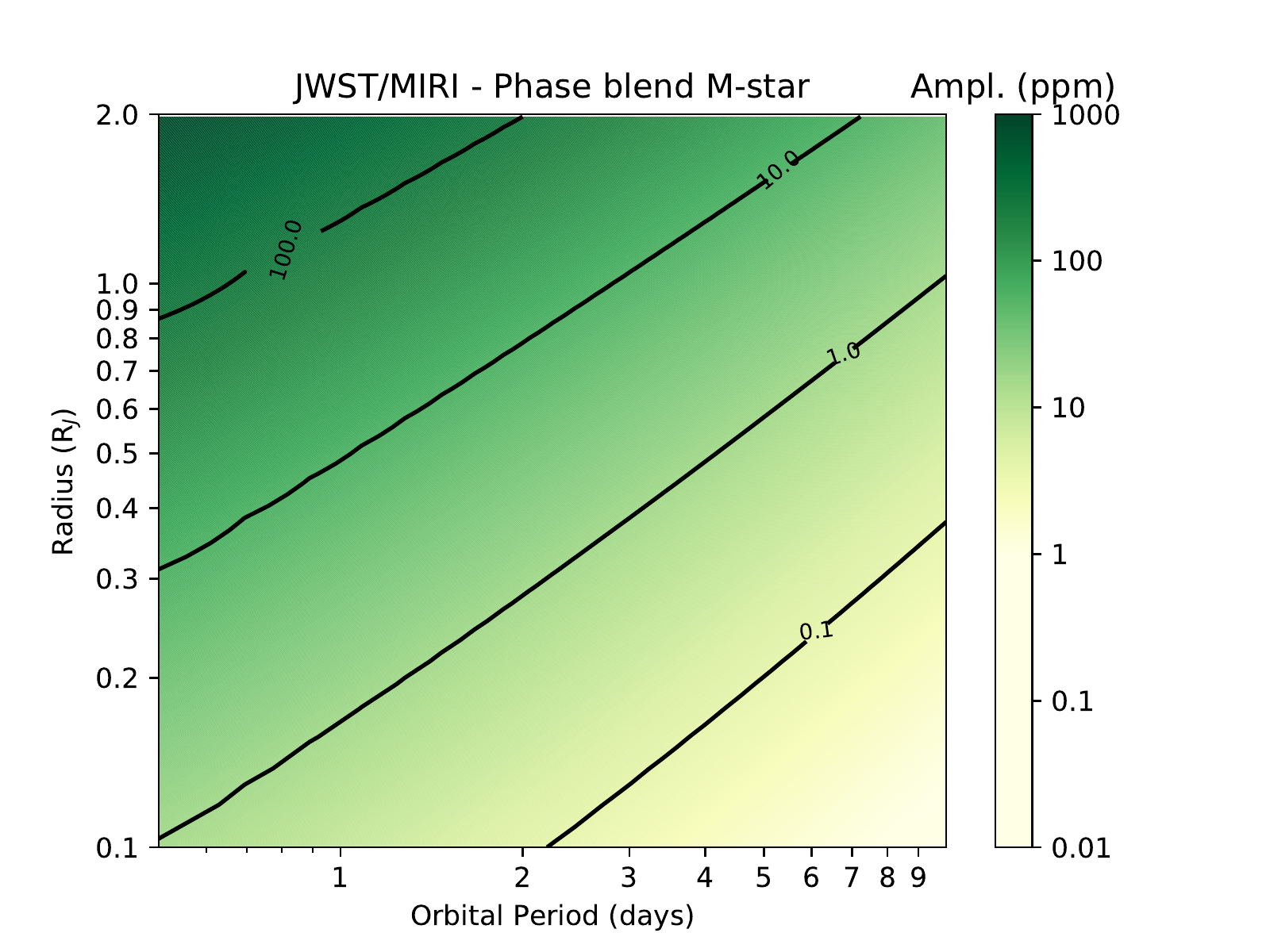}
\includegraphics[width=0.49\textwidth]{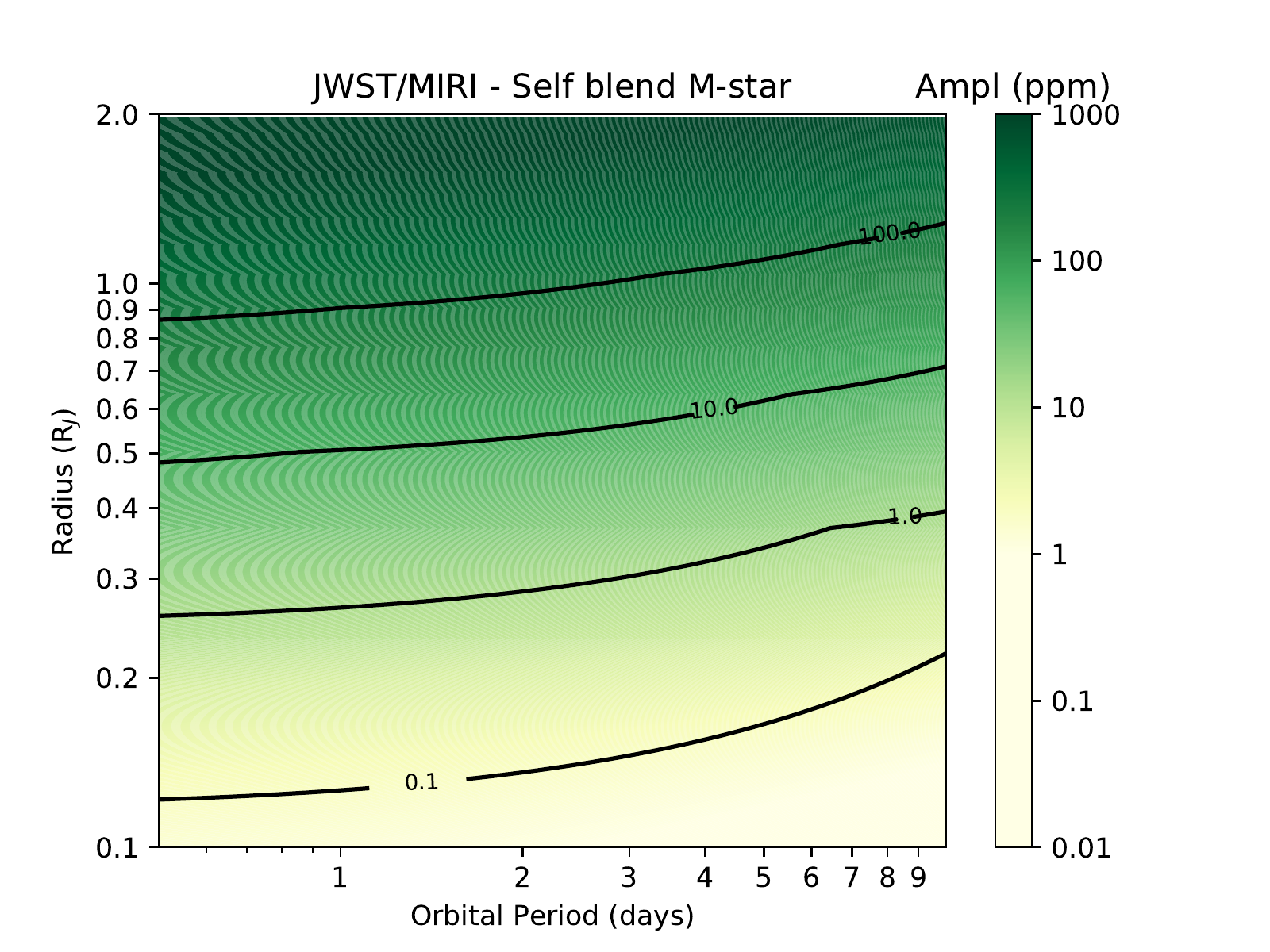}
\caption{Color maps analogous to those in Figure \ref{fig:colormap_Gstar_b0.5}, but for an M dwarf with $T_{*,\text{eff}} = 3800 \, K$, $M = 0.51 \, M_{\odot}$, and $R = 0.60 \, R_{\odot}$. The contour lines in the upper left panel appear to be unsmooth because of numerical precision at the $<$1 ppm level. \label{fig:colormap_Mstar_b0.5}}
\end{figure*}

\section{Conclusions} \label{sec:conclusions}
In this paper, we studied the impact of the common dark planet approximation in transit spectroscopy. We adopted the novel \texttt{ExoTETHyS.BOATS} subpackage to calculate the transit depth bias on simulated spectra with the JWST. This newly released software relies on the formulae derived by \cite{martin-lagarde2020} (the core algorithm was already used in that study). We find that, in some cases, these biases can significantly affect the parameters obtained from atmospheric retrievals, including gas temperature and mixing ratios. These effects are more important for large-radius, short-period planets. The color dependence appears to be stronger in near-infrared spectra and/or when combining UV/visible with mid-infrared spectra.
Our retrospective reanalysis for observations already planned within the JWST GTO and ERS programs did not reveal any critical issues, either because the bias is negligible or it should be detrendable from the data. We also provide precalculated tables and color maps as quick-look tools to assess the potential impact of self- and phase-blend biases on specific targets.
Knowing in advance an estimate of the bias can guide the selection of targets and/or the choice of the type of observation.

\acknowledgments
We thank the anonymous referee for their helpful suggestions. G.M. thanks Q. Changeat, C. Cossou, D. Dicken, and I. P. Waldmann for useful discussions.
This work was supported by the LabEx P2IO, the French ANR contract 05-BLANNT09-573739, and the European Union's Horizon 2020 Research and Innovation Programme (grant agreement No. 776403). T.Z. has been funded by the European Research Council (ERC) under the European Union's Horizon 2020 Research and Innovation Programme (grant agreement No. 679030/WHIPLASH).

\clearpage \startlongtable
\begin{deluxetable*}{ccccccc}%[t!]
\tablecaption{Transit depth biases and error bars estimated for the most likely configurations (see Table \ref{tab:T_Ab_eps}). The offset is the mean value of the bias over the wavelength bins. The amplitude is the difference between the maximum and minimum bias over the wavelength bins. Numbers in bold denote bias values greater than three times the nominal error bar, and greater than 30 ppm. (This table is available in machine-readable form.) \label{tab:spectral_biases}}
\tablecolumns{7}
%\tablenum{1}
\tablewidth{0pt}
\tablehead{
 & & \multicolumn2c{Total bias} & \multicolumn2c{Self-blend bias}\\
\colhead{Planet} & \colhead{Instrument} & \colhead{Offset} & \colhead{Amplitude} & \colhead{Offset} & \colhead{Amplitude} & \colhead{Error bar}
}
\startdata
 & NIRISS & \textbf{+170} & \textbf{242} & -4 & 9 & 42 \\
WASP~12~b & NIRSpec & \textbf{+275} & 119 & -12 & 17 & 57 \\
 & MIRI & +303 & 13 & -27 & 13 & 131 \\
\hline
 & NIRISS & +20 & 52 & -1 & 5 & 55 \\
WASP~43~b & NIRSpec & +58 & 58 & -10 & 25 & 65 \\
 & MIRI & +69 & 25 & -46 & 44 & 143 \\
\hline
 & NIRISS & \textbf{+157} & \textbf{241} & -4 & 9 & 29 \\
WASP~121~b & NIRSpec & \textbf{+266} & \textbf{127} & -13 & 19 & 40 \\
 & MIRI & \textbf{+298} & 14 & -32 & 18 & 93 \\
\hline
 & NIRISS & \textbf{+85} & \textbf{138} & -9e-1 & 1 & 19 \\
WASP~18~b & NIRSpec & \textbf{+159} & \textbf{108} & -2 & 1 & 26 \\
 & MIRI & \textbf{+217} & 28 & -3 & 2 & 60 \\
\hline
 & NIRISS & +2 & 3 & -1 & 5 & 23 \\
WASP~79~b & NIRSpec & +1 & 5 & -6 & 11 & 31 \\
 & MIRI & -6 & 7 & -16 & 8 & 73 \\
\hline
 & NIRISS & -2e-1 & 2 & -4e-1 & 3 & 26 \\
WASP~80~b & NIRSpec & -7 & 22 & -8 & 25 & 30 \\
 & MIRI & -48 & 58 & -53 & 60 & 64 \\
\hline
 & NIRISS & -6 & 18 & -6 & 18 & 33 \\
WASP~17~b & NIRSpec & -22 & 32 & -22 & 32 & 46 \\
 & MIRI & -47 & 20 & -47 & 20 & 106 \\
\hline
 & NIRISS & -1 & 7 & -1 & 7 & 47 \\
WASP~39~b & NIRSpec & -12 & 26 & -12 & 26 & 60 \\
 & MIRI & -42 & 31 & -42 & 31 & 137 \\
\hline
 & NIRISS & -4e-1 & 3 & -4e-1 & 3 & 16 \\
WASP~69~b & NIRSpec & -5 & 14 & -5 & 14 & 19 \\
 & MIRI & -25 & 22 & -25 & 22 & 43 \\
\hline
 & NIRISS & +2e-1 & 2 & -1 & 6 & 7 \\
HD~189733~b & NIRSpec & -6 & 19 & -11 & 27 & 9 \\
 & MIRI & -36 & 37 & -46 & 39 & 19 \\
\hline
 & NIRISS & +2 & 4 & -3e-1 & 1 & 7 \\
HD~209458~b & NIRSpec & +4 & 3 & -3 & 8 & 9 \\
 & MIRI & -2 & 9 & -13 & 12 & 21 \\
%\hline
\tablebreak
 & NIRISS & +1 & 3 & -1e-2 & 6e-2 & 9 \\
HD~149026~b & NIRSpec & +3 & 3 & -1e-1 & 2e-1 & 12 \\
 & MIRI & +5 & 4e-1 & -4e-1 & 3e-1 & 27 \\
\hline
 & NIRISS & +2e-1 & 6e-1 & -4e-1 & 2 & 25 \\
HAT-P-1 b & NIRSpec & -2 & 6 & -4 & 8 & 33 \\
 & MIRI & -10 & 9 & -14 & 10 & 76 \\
\hline
 & NIRISS & +3e-2 & 3e-2 & -4e-3 & 3e-2 & 39 \\
HAT-P-26 b & NIRSpec & -9e-2 & 4e-1 & -1e-1 & 4e-1 & 49 \\
 & MIRI & -1 & 1 & -1 & 1 & 110 \\
\hline
 & NIRISS & +1e-2 & 6e-2 & -4e-4 & 4e-3 & 13 \\
GJ~436~b & NIRSpec & +1e-1 & 3e-1 & -4e-2 & 2e-1 & 13 \\
 & MIRI & -4e-2 & 1 & -8e-1 & 2 & 27 \\
\hline
 & NIRISS & +2e-2 & 4e-3 & -4e-4 & 4e-3 & 24 \\
GJ~3470~b & NIRSpec & -1e-2 & 1e-1 & -3e-2 & 1e-1 & 26 \\
 & MIRI & -6e-1 & 1 & -6e-1 & 1 & 54 \\
\hline
\enddata
\end{deluxetable*}

\clearpage \startlongtable
\begin{deluxetable*}{ccccccc}%[t!]
\tablecaption{Transit depth biases and error bars estimated for the extreme configurations (see Table \ref{tab:T_Ab_eps}). (This table is available in machine-readable form.) \label{tab:spectral_biases_extreme}}
\tablecolumns{7}
%\tablenum{1}
\tablewidth{0pt}
\tablehead{
 & & \multicolumn2c{Total bias} & \multicolumn2c{Self-blend bias}\\
\colhead{Planet} & \colhead{Instrument} & \colhead{Offset} & \colhead{Amplitude} & \colhead{Offset} & \colhead{Amplitude} & \colhead{Error bar}
}
\startdata
 & NIRISS & \textbf{+257}/-19 & \textbf{354}/36 & -4/-19 & 6/36 & 42 \\
WASP~12~b & NIRSpec & \textbf{+423}/-41 & \textbf{229}/34 & -7/-41 & 4/34 & 57 \\
 & MIRI & \textbf{+549}/-61 & 75/13 & -9/-61 & 1/13 & 131 \\
\hline
 & NIRISS & +30/-10 & 80/34 & -9e-1/-10 & 2/34 & 55 \\
WASP~43~b & NIRSpec & +95/-47 & \textbf{118}/79 & -3/-47 & 3/79 & 65 \\
 & MIRI & +179/-117 & 61/59 & -5/-117 & 2/59 & 143 \\
\hline
 & NIRISS & \textbf{+171}/-21 & \textbf{271}/45 & -3/-21 & 5/45 & 29 \\
WASP~121~b & NIRSpec & \textbf{+312}/-51 & \textbf{203}/49 & -6/-51 & 4/49 & 40 \\
 & MIRI & \textbf{+428}/-82 & 71/21 & -8/-82 & 1/21 & 93 \\
\hline
 & NIRISS & \textbf{+98}/-7 & \textbf{152}/14 & -1/-7 & 2/14 & 19 \\
WASP~18~b & NIRSpec & \textbf{+175}/-16 & \textbf{110}/15 & -2/-16 & 1/15 & 26 \\
 & MIRI & \textbf{+238}/-26 & 38/6 & -3/-26 & 4e-1/6 & 60 \\
\hline
 & NIRISS & +7/-2 & 17/7 & -9e-2/-2 & 2e-1/7 & 23 \\
WASP~79~b & NIRSpec & +19/-9 & 21/13 & -2e-1/-9 & 3e-1/13 & 31 \\
 & MIRI & +34/-19 & 10/8 & -4e-1/-19 & 1e-1/8 & 73 \\
\hline
 & NIRISS & +7e-1/-6e-1 & 3/4 & -2e-2/-6e-1 & 1e-1/4 & 26 \\
WASP~80~b & NIRSpec & +6/-11 & 13/32 & -2e-1/-11 & 5e-1/32 & 30 \\
 & MIRI & +21/-62 & 16/65 & -7e-1/-62 & 6e-1/65 & 64 \\
\hline
 & NIRISS & +16/-6 & 36/19 & -3e-1/-6 & 7e-1/19 & 33 \\
WASP~17~b & NIRSpec & +41/-24 & 43/33 & -8e-1/-24 & 8e-1/33 & 46 \\
 & MIRI & +69/-49 & 19/20 & -1/-49 & 4e-1/20 & 106 \\
\hline
 & NIRISS & +1/-1 & 5/7 & -4e-2/-1 & 1e-1/7 & 47 \\
WASP~39~b & NIRSpec & +7/-12 & 13/27 & -2e-1/-12 & 3e-1/27 & 60 \\
 & MIRI & +19/-43 & 10/31 & -5e-1/-43 & 2e-1/31 & 137 \\
\hline
 & NIRISS & +5e-1/-5e-1 & 2/3 & -1e-2/-5e-1 & 4e-2/3 & 16 \\
WASP~69~b & NIRSpec & +3/-6 & 6/15 & -7e-2/-6 & 1e-1/15 & 19 \\
 & MIRI & +10/-26 & 6/23 & -2e-1/-26 & 1e-1/23 & 43 \\
\hline
 & NIRISS & +4/-3 & 12/14 & -1e-1/-3 & 3e-1/14 & 7 \\
HD~189733~b & NIRSpec & +16/-22 & 24/\textbf{44} & -4e-1/-22 & 7e-1/\textbf{44} & 9 \\
 & MIRI & +36/\textbf{-68} & 16/44 & -1/\textbf{-68} & 4e-1/44 & 19 \\
\hline
 & NIRISS & +4/-2 & 12/8 & -7e-2/-2 & 2e-1/8 & 7 \\
HD~209458~b & NIRSpec & +14/-11 & 19/20 & -2e-1/-11 & 3e-1/20 & 9 \\
 & MIRI & +28/-29 & 10/15 & -5e-1/-29 & 2e-1/15 & 21 \\
\hline
 & NIRISS & +2/-1e-1 & 5/4e-1 & -6e-3/-1e-1 & 1e-2/4e-1 & 9 \\
HD~149026~b & NIRSpec & +6/-5e-1 & 7/7e-1 & -2e-2/-5e-1 & 2e-2/7e-1 & 12 \\
 & MIRI & +11/-1 & 3/4e-1 & -3e-2/-1 & 8e-3/4e-1 & 27 \\
%\hline
\tablebreak
 & NIRISS & +1/-1 & 4/4 & -2e-2/-1 & 6e-2/4 & 25 \\
HAT-P-1 b & NIRSpec & 5/-6 & 7/12 & -7e-2/-6 & 1e-1/12 & 33 \\
 & MIRI & +11/-18 & 5/11 & -2e-1/-18 & 7e-2/11 & 76 \\
\hline
 & NIRISS & +2e-1/-6e-2 & 1/3e-1 & -2e-3/-6e-2 & 6e-3/3e-1 & 39 \\
HAT-P-26 b & NIRSpec & +1/-6e-1 & 3/2 & -9e-3/-6e-1 & 2e-2/2 & 49 \\
 & MIRI & +4/-3 & 2/2 & -2e-2/-3 & 1e-2/2 & 110 \\
\hline
 & NIRISS & +1e-2/-7e-3 & 9e-2/6e-2 & -1e-4/-7e-3 & 7e-4/6e-2 & 13 \\
GJ~436~b & NIRSpec & +2e-1/-3e-1 & 6e-1/9e-1 & -2e-3/-3e-1 & 5e-3/9e-1 & 13 \\
 & MIRI & +1/-2 & 1/3 & -9e-3/-2 & 9e-3/3 & 27 \\
\hline
 & NIRISS & +3e-2/-6e-3 & 2e-1/5e-2 & -2e-4/-6e-3 & 1e-3/5e-2 & 24 \\
GJ~3470~b & NIRSpec & +4e-1/-2e-1 & 1/6e-1 & -3e-3/-2e-1 & 7e-3/6e-1 & 26 \\
 & MIRI & +2/-1 & 2/2 & -1e-2/-1 & 1e-2/2 & 54 \\
\hline
%\enddata
%\end{deluxetable*}
%
%\begin{deluxetable*}{ccccccc}[t!]
%\tablecaption{Transit depth biases and error bars estimated for the extreme configurations (see Table \ref{tab:T_Ab_eps}). It continues from Table \ref{tab:spectral_biases_extreme}. \label{tab:spectral_biases_extreme2}}
%\tablecolumns{7}
%\tablenum{1}
%\tablewidth{0pt}
%\tablehead{
% & & \multicolumn2c{Total bias} & \multicolumn2c{Self-blend bias}\\
%\colhead{Planet} & \colhead{Instrument} & \colhead{Offset} & \colhead{Amplitude} & \colhead{Offset} & \colhead{Amplitude} & \colhead{Error bar}
%}
%\startdata
 & NIRISS & +35/-7 & \textbf{79}/21 & -7e-1/-7 & 2/21 & 24 \\
WASP~77A~b & NIRSpec & +90/-26 & \textbf{94}/38 & -2/-26 & 2/38 & 31 \\
 & MIRI & +152/-56 & 42/23 & -3/-56 & 8e-1/23 & 71 \\
\hline
 & NIRISS & +10/-6 & 30/26 & -3e-1/-6 & 9e-1/26 & 58 \\
WASP~52~b & NIRSpec & +36/-37 & 51/69 & -1/-37 & 2/69 & 73 \\
 & MIRI & +76/-104 & 31/60 & -2/-104 & 1/60 & 164 \\
\hline
 & NIRISS & +4/-1 & 12/5 & -6e-2/-1 & 2e-1/5 & 18 \\
WASP~127~b & NIRSpec & +14/-7 & 19/13 & -2e-1/-7 & 3e-1/13 & 24 \\
 & MIRI & +29/-19 & 11/10 & -4e-1/-19 & 1e-1/10 & 55 \\
\hline
 & NIRISS & +1e-1/-1e-1 & 6e-1/9e-1 & -2e-3/-1e-1 & 1e-2/9e-1 & 24 \\
WASP~107~b & NIRSpec & +1/-3 & 3/10 & -3e-2/-3 & 7e-2/10 & 29 \\
 & MIRI & +5/-21 & 5/26 & -1e-1/-21 & 1e-1/26 & 63 \\
\hline
 & NIRISS & +4/-2e-1 & 7/6e-1 & -9e-3/-2e-1 & 2e-2/6e-1 & 30 \\
TOI~193~b & NIRSpec & +8/-7e-1 & 6/8e-1 & -2e-2/-7e-1 & 1e-2/8e-1 & 39 \\
 & MIRI & +11/-1 & 2/4e-1 & -3e-2/-1 & 6e-3/4e-1 & 88 \\
\hline
 & NIRISS & +3e-2/-1e-2 & 2e-1/1e-1 & -5e-4/-1e-2 & 3e-3/1e-1 & 53 \\
GJ~1214~b & NIRSpec & +6e-1/-6e-1 & 2/2 & -1e-2/-6e-1 & 3e-2/2 & 49 \\
 & MIRI & +4/-8 & 5/12 & -7e-2/-8 & 8e-2/12 & 99 \\
\hline
 & NIRISS & +3e-4/-2e-5 & 2e-3/2e-4 & -3e-7/-2e-5 & 3e-6/2e-4 & 14 \\
GJ~357~b & NIRSpec & +9e-3/-2e-3 & 3e-2/7e-3 & -1e-5/-2e-3 & 4e-5/7e-3 & 15 \\
 & MIRI & +7e-2/-3e-2 & 1e-1/5e-2 & -9e-5/-3e-2 & 1e-4/5e-2 & 31 \\
\hline
 & NIRISS & +3e-3/-3e-4 & 3e-2/3e-3 & -1e-5/-3e-4 & 7e-5/3e-3 & 45 \\
GJ~1132~b & NIRSpec & +7e-2/-2e-2 & 2e-1/8e-2 & -2e-4/-2e-2 & 7e-4/8e-2 & 45 \\
 & MIRI & +5e-1/-2e-1 & 6e-1/4e-1 & -1e-3/-2e-1 & 2e-3/4e-1 & 92 \\
\hline
 & NIRISS & +3e-4/-3e-5 & 3e-3/3e-4 & -5e-7/-3e-5 & 5e-6/3e-4 & 19 \\
L98-59 c & NIRSpec & +1e-2/-3e-3 & 4e-2/1e-2 & -2e-5/-3e-3 & 7e-5/1e-2 & 19 \\
 & MIRI & +1e-1/-5e-2 & 1e-1/9e-2 & -2e-4/-5e-2 & 2e-4/9e-2 & 39 \\
%\hline
\tablebreak
 & NIRISS & +3e-6/-2e-6 & 4e-5/4e-5 & -7e-9/-2e-6 & 8e-8/4e-5 & 24 \\
L98-59 d & NIRSpec & +3e-4/-1e-3 & 1e-3/5e-3 & -8e-7/-1e-3 & 3e-6/5e-3 & 25 \\
 & MIRI & +5e-3/-4e-2 & 1e-2/9e-2 & -1e-5/-4e-2 & 2e-5/9e-2 & 51 \\
\hline
 & NIRISS & +1e-2/-1e-3 & 8e-2/1e-2 & -4e-5/-1e-3 & 3e-4/1e-2 & 163 \\
LP~791-18~b & NIRSpec & +2e-1/-5e-2 & 6e-1/2e-1 & -8e-4/-5e-2 & 2e-3/2e-1 & 150 \\
 & MIRI & +1/-6e-1 & 1/8e-1 & -5e-3/-6e-1 & 5e-3/8e-1 & 299 \\
\hline
 & NIRISS & +3e-4/-5e-5 & 4e-3/8e-4 & -3e-6/-5e-5 & 3e-5/8e-4 & 138 \\
Trappist-1 b & NIRSpec & +3e-2/-2e-2 & 1e-1/1e-1 & -2e-4/-2e-2 & 1e-3/1e-1 & 107 \\
 & MIRI & +4e-1/-6e-1 & 8e-1/2 & -4e-3/-6e-1 & 6e-3/2 & 202 \\
\hline
 & NIRISS & +5e-7/-6e-8 & 9e-6/1e-6 & -2e-9/-6e-8 & 4e-8/1e-6 & 119 \\
Trappist-1 d & NIRSpec & +3e-4/-2e-4 & 2e-3/2e-3 & -1e-6/-2e-4 & 8e-6/2e-3 & 92 \\
 & MIRI & +2e-2/-3e-2 & 4e-2/1e-1 & -6e-5/-3e-2 & 2e-4/1e-1 & 174 \\
\hline
 & NIRISS & +5e-8/-6e-9 & 9e-7/2e-7 & -2e-10/-6e-9 & 5e-9/2e-7 & 111 \\
Trappist-1 e & NIRSpec & +7e-5/-9e-5 & 5e-4/7e-4 & -4e-7/-9e-5 & 3e-6/7e-4 & 86 \\
 & MIRI & +6e-3/-3e-2 & 2e-2/1e-1 & -4e-5/-3e-2 & 1e-4/1e-1 & 163 \\
\hline
 & NIRISS & +2e-9/-4e-10 & 5e-8/1e-8 & -2e-11/-4e-10 & 4e-10/1e-8 & 105 \\
Trappist-1 f & NIRSpec & +1e-5/-2e-5 & 9e-5/2e-4 & -9e-8/-2e-5 & 7e-7/2e-4 & 82 \\
 & MIRI & +2e-3/-2e-2 & 8e-3/1e-1 & -2e-5/-2e-2 & 6e-5/1e-1 & 154 \\
\hline
\enddata
\end{deluxetable*}

\bibliography{myboats.bib}
\bibliographystyle{aasjournal}



\end{document}